\documentclass[cmp,final]{svjour}

\journalname{Communications in Mathematical Physics}

\renewcommand{\thesection}{\arabic{section}}
\makeatletter

\@addtoreset{equation}{section}
\makeatother

\def\ds{\displaystyle}
\def\eN{\varepsilon_N}

\def\e{\varepsilon}
\def\ov{\overline}
\def\ti{\tilde}
\def\d{\partial}
\def\la{\langle}
\def\ra{\rangle}
\def\lla{\left\langle}
\def\rra{\right\rangle}
\def\const{\,\mbox{const}\,}
\def\H{{\cal H}}
\def\fH{\mbox{H}}
\def\fA{\mbox{A}}
\def\bxm{\mbox{\boldmath$\xi^{(\mu)}$}}
\def\bxn{\mbox{\boldmath$\xi^{(\nu)}$}}
\def\bx1{\mbox{\boldmath$\xi^{(1)}$}}
\def\bxp{\mbox{\boldmath$\xi^{(p)}$}}
\def\bJ{\mbox{\boldmath$J$}}
\def\bh{\mbox{\boldmath$h$}}
\def\xmo{\xi^{(\mu)}_1}
\def\xmi{\xi^{(\mu)}_i}
\def\xni{\xi^{(\nu)}_i}
\def\xmj{\xi^{(\mu)}_j}
\def\xmN{\xi^{(\mu)}_N}

\def\xno{\xi^{(\nu)}_1}
\def\xm{x^{(\mu)}}

\def\tm{t^{(\mu)}}
\def\tn{t^{(\nu)}}
\def\vx{\ov x}
\def\no{\noindent}
\def\de{\delta}
\def\D{\Delta_N}
\def\P{\mbox{Prob}}

\def\a{\alpha}
\def\l{\lambda}

\begin{document}
\title{{Rigorous Solution of the Gardner Problem}}
\titlerunning{Rigorous Solution of the Gardner Problem}

\author{ Mariya Shcherbina\inst{1}\and
 Brunello Tirozzi\inst{2}}

\institute{Institute for Low
Temperature Physics, Ukr. Ac. Sci., 47 Lenin ave., Kharkov,
Ukraine, \email{shcherbi@ilt.kharkov.ua}\and Department of Physics of
 Rome University "La Sapienza", 5, p-za A.Moro, Rome, Italy,
\email{tirozzi@krishna.phys.uniroma1.it}}

\date{}

\maketitle

\begin{abstract}
We prove rigorously the well-known  result of Gardner about the typical
 fractional volume of interactions between $N$ spins which solve the 
 problem of storing a given set of $p$ random patterns. The Gardner formula
 for this volume  in the limit $N,p\to\infty$, $p/N\to\alpha$
 is proven for all values of $\alpha$. Besides, we prove a useful criterion
 of the factorisation of all correlation functions for a class of
 spin glass model.
 
\end{abstract}

\section{Introduction}

The spin glass and neural network theories are of considerable importance 
and interest for a number of 
branches of theoretical and mathematical
physics (see \cite{MPV} and references therein). Among many topics of 
interest  the analysis of the different models of neural network dynamics
is one of the most important.
 
The neural network dynamics is defined as 
\begin{equation}
\sigma_i(t+1)=\hbox{sign}\{\sum_{j=1,j\not=i}^N  J_{ij}\sigma_j(t)\}
\quad (i=1,\dots, N),
\label{dyn}\end{equation}
where $\{\sigma_j(t)\}_{j=1}^N$ are the Ising spins and the interaction matrix
$\{J_{ij}\}$ (not necessarily symmetric) depends on the concrete model, but
usually it satisfies the conditions
\begin{equation}
\sum_{j=1,j\not=i}^N J_{ij}^2=NR\quad (i=1,\dots, N),
\label{cond_J}\end{equation}
where $R$ is some fixed number which could be taken equal to 1.

The main problem of the neural network theory is to 
introduce an interaction in such a way that   some chosen vectors 
$\{\bxm\}_{\mu=1}^p$ (patterns) are the fixed points of the dynamics 
(\ref{dyn}).
 This implies the conditions:
\begin{equation}
\xmi\sum_{j=1,j\not=i}^N  J_{ij}\xmj>0\quad (i=1,\dots, N).
\label{fp}\end{equation}
Usually, to simplify the problem the patterns 
$\{\bxm\}_{\mu=1}^p$ are chosen i.i.d. random vectors with 
i.i.d. components 
$\xmi$ ($i=1,\dots,N$), assuming values $\pm 1$ with probability $\frac{1}{2}$.

Sometimes condition (\ref{fp}) is not sufficient to have $\bxm$ as
the end points of the dynamics. To have some "basin of attraction" (that is
some neighbourhood of $\bxm$, starting from which we for sure arrive in
$\bxm$) one should introduce some positive parameter $k$ and 
impose the conditions: 
\begin{equation}
\xmi\sum_{j=1,j\not=i}^N \ti J_{ij}\xmj>k \quad (i=1,\dots, N).
\label{fp1}\end{equation}
Gardner \cite{G} was the first who solved a kind of inverse
problem. She asked the questions: for which 
$\alpha=\frac{p}{N}$ the interaction $\{J_{ij}\}$, satisfying (\ref{cond_J}) and (\ref{fp1})
exists? What is the typical fractional volume of these interactions? 
Since all  condition (\ref{cond_J})
and (\ref{fp1}) are factorised with respect to $i$, this problem
 after a simple transformation should be replaced by the following.
For the system of $p\sim\alpha N$ i.i.d. random  patterns
$\{\bxm\}_{\mu=1}^p$ with i.i.d. $\xmi$ ($i=1,\dots,N$)
assuming values $\pm 1$ with probability $\frac{1}{2}$, consider
\begin{equation}
\Theta_{N,p}(k)=\sigma_N^{-1}\int_{(\bJ,\bJ)=N} d\bJ\prod_{\mu=1}^p
\theta(N^{-1/2}(\bxm,\bJ)-k),
\label{Theta}\end{equation}
where the function $\theta(x)$, as usually, is zero in the negative semi-axis and 1
in the positive and  $\sigma_N$ is the Lebesgue measure of $N$-dimensional 
sphere of radius $N^{1/2}$. Then, the question of interest is the behaviour of
$\frac{1}{N}\log\Theta_{N,p}(k)$ in the limit $N,p\to \infty$, 
$\frac{p}{N}\to\alpha$.
Gardner \cite{G} had solved this problem  by using the
so-called replica trick, which is completely non-rigorous
from the  mathematical point of view but sometimes very useful in the 
physics of spin glasses (see \cite{MPV} and references therein). She obtained 
that for any $\alpha<\alpha_c(k)$, where 
\begin{equation}
\a_c(k)\equiv ({1\over \sqrt{2\pi}}\int_{-k}^\infty(u+k)^2e^{-u^2/2}du)^{-1},
\label{a_c}\end{equation}
there exists
\begin{equation}\begin{array}{c}\ds{
\lim_{N,p\to\infty,p/N\to\a}E\{\log \Theta_{N,p}(k)\}={\cal F}(\alpha,k)}\\
\ds{
\equiv\min_{q:0\le q\le 1}\left[\a
E\left\{\log \emph{H}\left({u\sqrt q+k\over\sqrt{1-q}}\right)\right\}\right.
\left.+{1\over 2}{q\over 1-q}+{1\over 2}\log(1-q)\right],}
\end{array}\label{cal-F}\end{equation}
where $u$ is the Gaussian random variable with zero mean and variance $1$, and
here and below we denote by the symbol $E\{...\}$ the averaging with respect
to all random parameters of the problem. And 
$\frac{1}{N}\log\Theta_{N,p}(k)$ tends to minus infinity for 
$\alpha\ge\alpha_c(k)$.

\smallskip

At the present paper we give the rigorous mathematical proof of the Gardner
results. As far as we know, it is one of the first cases, when the problem of
spin glass theory can be  completely (i.e. for all parameters $\alpha$ and
$k$) solved in the rigorous mathematical way. It can be explained by the 
fact, that in the Gardner problem
the so-called replica symmetry solution is true for all  $\alpha$ and
$k$, while, e.g. in the Hopfield and Sherrington-Kirkpatrick models the
replica symmetry solution is valid only for small enough $\alpha$ or
for  high temperatures (see \cite{MPV} for the physical theory
and \cite{S}, \cite{S2},  \cite{Tal1}, \cite{Tal2}
for the respective rigorous results). The same situation holds, unfortunately, 
with a problem similar to the Gardner one, the so-called Gardner-Derrida 
\cite{GD} problem. Also only the case of small enough $\alpha$ was studied 
rigorously for this model (see \cite{Tal3}).

We solve the Gardner problem in three steps which are Theorems \ref{thm:1},
\ref{thm:2} and \ref{thm:3} below. At the first step we prove
 some general statement. We study an abstract situation, where
 the energy function (the Hamiltonian) and the configuration
space are convex and prove that in this case  all the correlation functions 
become
factorised in the thermodynamic limit. Usually this factorisation means that the 
ground state and the Gibbs measure are uniquely defined. In fact, physicists 
understood this fact during a rather long time, but in the rigorous 
mathematical way it was not proved before.

The proof of Theorem \ref{thm:1} is based on the application 
 of the theorem of classical geometry, known since the nineteenth
century as the Brunn-Minkowski theorem. This theorem studies the
intersections of a convex set with the family of parallel
hyper-planes (see the proof of Theorem \ref{thm:1}
for the exact statement). We only need to prove some corollary from this
theorem (Proposition \ref{pro:1}), which allows us to have $N$- independent 
estimates. As a result we obtain the rigorous proof of the general factorisation
property of all correlation functions (see (\ref{t1.1})).
Everybody who is familiar with the theory of spin glasses knows that
decay of correlations is the key point in the derivation of self-consistent
equations.

  \smallskip
 
 The second  step is the derivation of  self-consistent equations
 for the order parameters of our model. In fact Theorem \ref{thm:1}
 provides all the necessary to express the free energy in  terms of
 the order parameters, but the problem is that we are not able to
 produce the equations for these parameters in the case, when the
 "randomness" is not included in the Hamiltonian, but is connected with
  the  integration domain. That is why we use a rather common trick in 
 mathematics: substitute  $\theta$-functions by some  smooth
 functions which depend on the small parameter $\e$ and tend, 
 as $\e\to 0$, to $\theta$-function. We choose for these purposes
 $\fH(x\e^{-1/2})$, where $\fH$ is the $erf$-function (see definition
 (\ref{H})). But the particular form of these smoothing functions
 is not very important for us. The most important fact is, that they are not
 zero in any point and so, taking their logarithms, we can treat them
 as a part of our Hamiltonian.
 
 The proof of Theorem \ref{thm:2} is based on the the application  
to the Gardner problem
of the so-called cavity method, the rigorous version of which was
proposed in \cite{PS} and developed in 
\cite{S}, \cite{PST1}, \cite{PST2}. But in the previous papers (\cite{PS},\cite{PST1}, 
\cite{PST2}) we assumed the
factorisation of the correlation functions in the thermodynamic limit 
and on the basis of this fact derived the replica symmetry equation for the order 
parameters (to be more precise, we assumed that the order parameter possesses
the self-averaging property and obtained from this fact the factorisation
of the correlation function). Here, due to Theorem \ref{thm:1}, we can
prove the asymptotical factorisation property, 
which allows us to finish completely the study of the Gardner model.

\smallskip

Our last  step is the limiting transition $\e\to 0$, i.e. the proof that
the product of $\alpha N$ $\theta$-functions in (\ref{Theta}) can be replaced by 
the product of $\fH(\frac{x}{\sqrt{\e}})$ with the  small difference,
when $\e$ is small enough. Despite  our
expectations, it is the most difficult step from the technical point of view. 
It is rather simple to prove, that the expression (\ref{cal-F}) is an upper bound
 or $\log\Theta_N,p(k)$. But the estimate from below is much more complicated.
The problem  is that to estimate the difference between 
the free energies corresponding to two  Hamiltonians
we, as a rule, need to have them defined in the common configuration space, 
or, at least, we need to  know some a priori bounds for some Gibbs
averages. In the case of the Gardner problem we do not possess this
information. This leads to rather serious (from our point of
view) technical problems
(see the proof of Theorem \ref{thm:3} and Lemma \ref{lem:4}).

The paper is organised as follows. The main definitions and results are
formulated in Sec.2. The proof of these results are given in Sec.3.
The auxiliary results (lemmas and propositions) which we need for the
proof are formulated in the text of Sec.3 and their proofs are
given in Sec.4.

\section{Main Results}

As it was mentioned above, we start from the abstract statement,
which allows us to prove the factorisation of all correlation
functions for some class of  models.

Let $\{\Phi_N(\bJ)\}_{N=1}^\infty$
($\bJ\in{\bf R}^N$) be a system of  convex functions
which possess the third derivatives, bounded in any
compact. Consider also a system of convex domains
$\{\Gamma_N\}_{N=1}^\infty$
($\Gamma_N\subset{\bf R}^N$) whose boundaries consist of a
finite number (may be depending on $N$) of smooth pieces.
We remark here, that for the Gardner problem we need to study $\Gamma_N$
which is the intersection of $\alpha N$ half-spaces 
but in Theorem \ref{thm:1} (see below) we consider a more general sequence
 of convex sets.
Define the  Gibbs measure and the free energy, corresponding to
$\Phi_N(\bJ)$ in $\Gamma_N$:
\begin{equation}\begin{array}{c}
\la\dots\ra_{\Phi_N}\equiv \Sigma_N^{-1}\int_{\Gamma_N}d\bJ(\dots)
\exp\{-\Phi_N(\bJ)\},\\
 \Sigma_N(\Phi_N)\equiv\int_{\Gamma_N}d\bJ
\exp\{-\Phi_N(\bJ)\},\,\,\, f_N(\Phi_N)\equiv{1\over N}\log \Sigma_N(\Phi_N).
\end{array}\label{gen-Z}\end{equation}
Denote
\begin{equation}\begin{array}{c}
\tilde\Omega_N(U)\equiv\{\bJ:\Phi_N(\bJ)\le NU\},\quad
\Omega_N (U)\equiv \tilde\Omega_N(U)\cap\Gamma_N,
\\
{\cal D}_N(U)\equiv\tilde{\cal D}_N(U)\cap \Gamma_N ,
\end{array}\label{Omega} \end{equation}
 where $\tilde{\cal D}_N(U)$ is the boundary of $\tilde\Omega_N(U)$. Then
define
$$
f_N^*(U)={1\over N}\log \int_{J\in {\cal D}_N(U)}d\bJ e^{-NU}.
$$

\begin{theorem}\label{thm:1}
 Let the functions $\Phi_N(\bJ)$ satisfy the conditions:
\begin{equation}
\frac{d^2}{dt^2}\Phi_N(\bJ+t{\bf e})|_{t=0}\ge C_0>0,
\label{cond1}\end{equation}
with any direction ${\bf e}\in{\bf R}^N,\,\,|{\bf e}|=1$ and
uniformly in any set $|\bJ|\le N^{1/2} R_1$,
\begin{equation}
\Phi_N(\bJ)\ge C_1(\bJ,\bJ), \quad as \,\,\, (\bJ,\bJ)>NR^2,
\label{cond2}\end{equation}
and for any $\ds{U>U_{min}\equiv \min_{J\in \Gamma_N}N^{-1}\Phi_N(\bJ)\equiv 
N^{-1}\Phi_N(\bJ^*)}$
\begin{equation}
|\nabla \Phi_N(\bJ)|\le N^{1/2} C_2(U),\quad as \quad \bJ\in
\tilde\Omega_N(U)
\label{cond3} \end{equation}
with some positive $N$-independent $C_0,C_1,C_2(U)$  and 
$C_2(U)$ continuous in $U$.

Assume also, that there exists some finite $N$-independent $C_3$, such that
\begin{equation}
f_N(\Phi_N)\ge -C_3.
\label{cond4} \end{equation}
Then
\begin{equation}
|f_N(\Phi_N)-f_N^*(U_*)|\le O({\log N\over N}),\quad
\left( U_*\equiv\frac{1}{N}\la\Phi_N\ra_{\Phi_N}\right).
\label{t1.2}\end{equation}
Moreover, for any ${\bf e}\in{\bf R}^N$ $(|{\bf e}|=1)$ and any natural $p$
\begin{equation}
\la(\dot{\bJ},{\bf e})^p\ra_{\Phi_N}\le C(p) \quad (\dot J_i\equiv
J_i-\la J_i \ra_{\Phi_N})
\label{t1.1} \end{equation}
 with some positive $N$-independent $C(p)$.
\end{theorem}

Let us remark  that the main conditions here are, of course, the condition
that the domain $\Gamma_N$ and the Hamiltonian $\Phi_N$ are convex 
(\ref{cond1}). Condition (\ref{cond2}) and (\ref{cond3}) are not very 
restrictive, because they are fulfilled for the most part of Hamiltonians.
The bound (\ref{cond4}) in fact is the condition on the domain
$\Gamma_N$. This condition prevents $\Gamma_N$ to be too small. In the
application to the Gardner problem the existence of such a bound is very 
important, because in this case we should study just the question of 
the measure of $\Gamma_N$, which is the intersection of $\alpha N$ random
half-spaces with the  sphere of radius $N^{1/2}$. But
from the technical point of view for us it is more convenient to check
the existence of the bound from below for the free energy, than for
the volume of the configuration space (see the proof of Theorem \ref{thm:3}
below).

Theorem \ref{thm:1} has two rather important for us corollaries.

\begin{corollary}\label{cor:1} Under conditions
(\ref{cond1})- (\ref{cond4}) for any $U>U_{min}$
\begin{equation}
f_N^*(U)=\min_{z>0}\left\{f_N(z\Phi_N)+zU\right\}+O({\log N\over N}).
\label{c.1}\end{equation}
\end{corollary}

This  corollary is a simple generalisation of the so called spherical model
which becomes rather popular in the resent time  (see, e.g.  the review paper 
\cite{KKPS} and references therein). It allows us to substitute the integration
over the level surface of the function $\Phi_N$ by the integration
over the whole space, i.e. to substitute the "hard condition"
$\Phi_N=UN$ by the "soft one" $\la\Phi_N\ra_{\Phi_N}=UN$. It is a common trick
which  often  is very useful in statistical mechanics.

The second corollary gives the most important and convenient  form of
 the general property (\ref{t1.1}):

\begin{corollary}\label{cor:2}
Relations (\ref{t1.1}) imply that uniformly in N
$$
{1\over N^2}\sum\la\dot J_i \dot J_j\ra_{\Phi_N}^2\le{C\over N}.
$$
\end{corollary}

\smallskip

To found the free energy of the model (\ref{Theta}) and to derive the
replica symmetry equations for the order parameters we introduce
the "regularised" Hamiltonian, depending on the small parameter
$\e>0$
 
\begin{equation}
\H_{N,p}(\bJ,k,h,z,\e)\equiv-\sum_{\mu=1}^p\log \fH\left(
\frac{k-(\bxm,\bJ)N^{-1/2}}{\sqrt\e}\right)+h(\bh,\bJ)+
{z\over 2}(\bJ,\bJ),
\label{H_N,p}\end{equation}
where   the function $\fH(x)$ is defined as
\begin{equation}
\fH(x)\equiv{1\over\sqrt{2\pi}}\int_x^\infty e^{-t^2/2}dt
\label{H}\end{equation}
and $\bh=(h_1,...,h_N)$ is an external random
field with independent Gaussian $h_i$ with zero mean
and variance $1$, which we need from the technical reasons.

The partition function for this Hamiltonian is
\begin{equation}
Z_{N,p}(k,h,z,\e)=\sigma_N^{-1}\int d\bJ\exp\{-\H_\e(\bJ,h,z,\e)\}.
\label{Z}\end{equation}
We denote also by $\la\dots\ra$ the corresponding Gibbs averaging and
\begin{equation}
f_{N,p}(k,h,z,\e)\equiv{1\over N}\log Z_{N,p}(k,h,z,\e).
\label{f}\end{equation}

\begin{theorem}\label{thm:2}
For any $\a,k\ge 0$  and $z>0$  the functions $f_{N,p}(k,h,z,\e)$ are
self-averaging in the limit $N,p\to\infty$, $\a_N\equiv{p\over N}\to\a$:
\begin{equation}
E\left\{(f_{N,p}(k,h,z,\e)-E\{f_{N,p}(k,h,z,\e)\})^2\right\}\to 0
\label{s-a}\end{equation}
and, if $\e$ is small enough, $\a<2$ and $z\le \e^{-1/3}$, then there exists
\begin{equation}\begin{array}{c}\ds{
\lim_{N,p\to\infty,\a_N\to\a}E\{f_{N,p}(k,h,z,\e)\}=
F(\a,k,h,z,\e),}\\
\ds{
F(\a,k,h,z,\e)\equiv\max_{ R>0}\min_{0\le q\le R}\left[\a
E\left\{\log \emph{H}\left({u\sqrt q+k\over\sqrt{\e+R-q}}\right)\right\}
\right.}\\
\ds{ 
\left.+{1\over 2}{q\over R-q}+{1\over 2}\log(R-q)-{z\over 2}R
+{h^2\over 2}( R-q)\right],}
\end{array}\label{t2.1}\end{equation}
where  $u$ is a Gaussian random variable with zero mean and variance 1.
\end{theorem}

Let us note that the bound $\alpha<2$ is not important for us, because
for any $\alpha>\alpha_c(k)$ ($\alpha_c(k)$ is defined by (\ref{a_c}) and
$\alpha_c(k)<2$ for any $k$)
the free energy of the problem (\ref{Theta}) tends to
$-\infty$, as $N\to \infty$ (see Theorem \ref{thm:3} for the exact
statement). The bound $z<\e^{-1/3}$ also is  not a restriction for us.
We could need to consider $z>\e^{-1/3}$ only if, applying
(\ref{c.1}) to the Hamiltonian (\ref{H_N,p}), we obtain  that
 the point of minimum $z_{min}(\e)$ in (\ref{c.1}) does not satisfy this bound.
But it is shown in Theorem \ref{thm:3}, that for any $\alpha <\alpha_c(k)$
$z_{min}(\e)< \ov z$ with some  finite $\ov z$ depending only on $k$ and $\alpha$.

\smallskip

We start the analysis of  $\Theta_{N,p}(k)$, defined in (\ref{Theta}),
from the following remark.

\begin{remark}\label{rem:1}
Let us note that $\Theta_{N,p}(k)$ can be
zero with nonzero probability (e.g., if for some $\mu\not=\nu$
$\bxm=-\bxn$). Therefore we cannot,  as usually,
just take $\log \Theta_{N,p}(k)$.
To avoid this difficulty, we take some large enough $M$ and replace below
the $\log$- function by the function $\log_{(MN)}$, defined as
\begin{equation}
\log_{(MN)}X=\log\max\left\{X,\,\,\, e^{-MN}\right\}.
\label{log_M}\end{equation}
\end{remark}

\begin{theorem}\label{thm:3} 
For any $\a\le\a_c(k)$  $N^{-1}\log_{(MN)}\Theta_{N,p}(k)$
is self-averaging in the limit $N,p\to\infty$, $p/N\to\a$
$$
E\left\{\left(N^{-1}\log_{(MN)}\Theta_{N,p}(k)-
E\{ N^{-1}\log_{(MN)}\Theta_{N,p}(k)\}\right)^2\right\}\to 0
$$
and for $M$ large enough there exists
\begin{equation}\begin{array}{c}\ds{
\lim_{N,p\to\infty,p/N\to\a}E\{ N^{-1}\log_{(MN)}\Theta_{N,p}(k)\}=
{\cal F}(\alpha, k), }
\end{array}\label{t3.1}\end{equation}
where ${\cal F}(\alpha, k)$ is defined by (\ref{cal-F}).

For $\a>\a_c(k)$ $E\{N^{-1}\log_{(MN)}\Theta_{N,p}(k)\}\to -\infty$, as
$N\to\infty$ and then $M\to\infty$.
\end{theorem}
We would like to mention here that the self-averaging of 
$N^{-1}\log \Theta_{N,p}(k)$ was proven in (\cite{Tal4}), but 
 our proof of this fact is  necessary for the proof of (\ref{t3.1}).

\section{Proof of the Main Results}

\no{\it Proof of Theorem \ref{thm:1}}
For any $U>0$ consider the set $\Omega_N (U)$  defined in (\ref{Omega})
Since $\Phi_N(\bJ)$ is a convex function, the set $\Omega_N (U)$ is also
convex and $\Omega_N (U)\subset \Omega_N (U')$, if $U<U'$.
Let
\begin{equation}\begin{array}{c}
V_N(U)\equiv\hbox{mes}(\Omega_N (U)),\quad
S_N(U)\equiv\hbox{mes}({\cal D}_N(U)),\\
F_N(U)\equiv\int_{\bJ\in{\cal D}_N(U)}|\nabla\Phi_N(\bJ)|^{-1}dS_J.
\end{array}\label{t1.3}\end{equation}
Here and below  the symbol $\hbox{mes}(...)$ means the Lebesgue
measure of the correspondent dimension.

Then it is easy to see that the partition function $\Sigma_N$ can be
represented in the form
\begin{equation}\begin{array}{c}\ds{
\Sigma_N=\int_{U>U_{min}}e^{-NU}F_N(U)dU=
N^{-1}\int_{U>U_{min}}e^{-NU}{d\over dU}V_N(U)dU}\\
\ds{
=\int_{U>U_{min}}e^{-NU}V_N(U)dU}.
\end{array}\label{t1.4}\end{equation}
Here we have used the relation $F_N(U)=N^{-1}{d\over dU}V_N(U)$ and the integration
by parts.

Besides, for a chosen direction ${\bf e}\in{\bf R}^N$ ($|\bf e|=1$),
 and any real $c$  consider the hyper-plane
$$
{\cal A}(c,{\bf e})=\left\{\bJ\in {\bf R}^N: ({\bJ},{\bf e})=N^{1/2} c\right\}
$$
and denote
\begin{equation}\begin{array}{c}
\ds{
\Omega_N (U,c)\equiv\Omega_N(U)\cap {\cal A}(c,{\bf e}),\quad
V_{N}(U,c)\equiv\hbox{mes}(\Omega_N (U,c)),}\\
\ds{
{\cal D}_N(U,c)\equiv{\cal D}_N(U)\cap {\cal A}(c,{\bf e}),
\quad F_N(U,c)\equiv\int_{\bJ\in{\cal D}_N(U,c)}|\nabla\Phi_N(\bJ)|^{-1}dS_J}.
\end{array}\label{t1.5}\end{equation}
 Then, since $F_N(U,c)=
N^{-1}{\d\over \d U}V_{N}(U,c)$, we obtain
\begin{equation}\begin{array}{c}\ds{
\Sigma_N=\int dcdUe^{-NU}F_N(U,c)=
\int dcdUe^{-NU}V_N(U,c)},\\
\ds{
\lla (\bJ,{\bf e})^p\rra_{\Phi_N}=
\frac{N^{p/2}\int dc dUc^pe^{-NU}V_N(U,c)}
{\int dc dUe^{-NU}V_N(U,c)}.}
\end{array}\label{t1.6}\end{equation}
Denote
\begin{equation}
s_N(U)\equiv \frac{1}{N}\log V_N(U),\quad
s_N(U,c)\equiv \frac{1}{N}\log V_N(U,c).
\label{t1.7}\end{equation}
Then relations  (\ref{t1.4}), (\ref{t1.6}) give us
\begin{equation}\begin{array}{c}\ds{
\Sigma_N=N\int\exp\{N(s_N(U)-U)\}dU,}\\
\ds{
\lla(\dot{\bJ},{\bf e})^p\rra_{\Phi_N}=
N^{p/2}\lla(c-\la c\ra_{(U,c)})^p\rra_{(U,c)},}
\end{array}\label{t1.8}\end{equation}
where
\begin{equation}\begin{array}{c}\ds{
\la ...\ra_{(U,c)}\equiv\frac{\int dU dc (...)\exp\{N(s_N(U,c)-U)\}}
{\int dU dc \exp\{N(s_N(U,c)-U)\}}.}
\end{array}\label{t1.8a}\end{equation}
Then (\ref{t1.2}) and (\ref{t1.1}) can be obtained by the standard
Laplace method, if we prove that $ s_N(U)$ and $ s_N(U,c)$ are
 concave functions  and they are strictly concave
in the neighbourhood of the   points of  maximum
of the functions  $(s_N(U)-U)$ and $(s_N(U,c)-U)$.
To prove this  we apply
the theorem of Brunn-Minkowski  from classical geometry
(see e.g. \cite{Had}) to the functions $s_N(U)$ and $s_N(U,c)$.
To formulate this theorem we need some extra
definitions.
\begin{definition}
Consider two bounded  sets in ${\cal A},{\cal B}\subset{\bf R}^N$.
For any positive $\alpha$ and $\beta$ 
$$
\alpha{\cal A}\times\beta{\cal B}\equiv
\left\{{\bf s}: {\bf s}=\alpha{\bf a}+\beta{\bf b}, {\bf a}\in{\cal A},
{\bf b}\in{\cal B}\right\}.
$$
$\alpha{\cal A}\times\beta{\cal B}$ is the Minkowski sum of $\alpha{\cal A}$
and $\beta{\cal B}$.
\end{definition}
\begin{definition}
The one-parameter family of bounded sets
$\{{\cal A}(t)\}_{t_1^*\le t\le t_2^*}$ is  a  convex
one- parameter family, if
for any positive $\alpha<1$ and $t_{1,2}\in [t_1^*,t_2^*]$
they satisfy the condition
$$
{\cal A}( \alpha t_1+(1-\alpha)t_2) \supset\alpha{\cal A}(t_1)\times
(1-\alpha){\cal A}(t_2).
$$
\end{definition}

{\bf Theorem of Brunn-Minkowski}
{\it Let $\{{\cal A}(t)\}_{t_1^*\le t\le t_2^*}$
be some convex one-parameter family. Consider
$R(t)\equiv(\emph{mes}{\cal A}(t))^{1/N}$. Then ${d^2R(t)\over dt^2}\le 0$
and ${d^2R(t)\over dt^2}\equiv 0$ for $t\in[t_1',t_2']$
if and only if
all the sets ${\cal A}(t)$ for $t\in[t_1',t_2']$  are homothetic to
each other.}

\medskip

For the proof of this theorem see, e.g., \cite{Had}.

\medskip
To use this theorem for the proof of (\ref{t1.2})
let us observe that the family
$\{\Omega_N(U))\}_{U>U_{min}}$ is a convex
one-parameter family and then, according to the Brunn-Minkowski theorem,
the function $R(U)=(V_N(U))^{1/N}$ is a concave function.
Thus, we get that $s_N(U)$ is a concave function:
$$
{d^2\over dU^2}s_N(U)={d^2\over dU^2}\log R(U)={R''(U)\over R(U)}-
\left({R'(U)\over R(U)}\right)^2\le -\left({R'(U)\over R(U)}\right)^2.
$$
 But  ${R'(U)\over R(U)}=\frac{d}{dU}s_N(U) >1$ for $U<U^*$, and even
 if $\frac{d}{dU}s_N(U)=0$ for $U>U^*$, we obtain that 
 $\frac{d}{dU}(s_N(U)-U)=-1$. Thus, using the standard Laplace method, we get
\begin{equation}\begin{array}{c}\ds{
f_N(\Phi_N)=s_N(U^*)-U^*+O({\log N\over N})=\frac{1}{N}\log V_N(U^*)
-U^*+O(\frac{\log N}{N}),}\\
U_*\equiv\frac{1}{N}\la \Phi_N \ra_{\Phi_N}=U^*+o(1).
\end{array}\label{t1.9}\end{equation}

 Using condition (\ref{cond3}), and taking $\bJ^*$, which is the minimum
 point of $\Phi_N(\bJ)$, we get
\begin{equation}\begin{array}{c}\ds{
V_N(U^*)\ge N^{-1}\int_{\bJ\in {\cal D}_N(U^*)}
|(\bJ-\bJ^*,\nabla\Phi_N(\bJ))| |\nabla\Phi_N(\bJ)|^{-1}
dS_{\bJ}}\\
\ds{
\ge S_N(U^*)\frac{U^*-U_{min}}
{\max_{\bJ\in {\cal D}_N(U^*)}|\nabla\Phi_N(\bJ)|} =
N^{-1/2}S_N(U^*)C(U^*)}.
\end{array}\label{t1.9a}\end{equation}
On the other hand, for any $U<U^*$
\begin{equation}\begin{array}{c}\ds{
\frac{S_N(U)}{N^{1/2} V_N(U)}\ge
\min_{\bJ\in{\cal D}_N(U)}|\nabla\Phi_N(\bJ)|\frac{F_N(U)}{N^{1/2} V_N(U)}}\\
\ds{
\ge N^{1/2}\min_{\bJ\in{\cal D}_N(U)}\frac{U-U_{min}}{|\bJ-\bJ^*|}
\frac{d}{dU}s_N(U)\ge \ti C \frac{d}{dU}s_N(U)>\ti C.}
\end{array}\label{t1.10}\end{equation}
Here we have used (\ref{t1.5}) and (\ref{cond2}). Thus the same inequality 
is valid also for $U=U^*$. Inequalities (\ref{t1.10}) and (\ref{t1.9a})
 imply that
$$
\frac{1}{N}\log S_N(U^*)=\frac{1}{N}\log V_N(U^*)+O(\frac{\log N}{N}).
$$
Combining this relation with (\ref{t1.9}) we get (\ref{t1.2}).

Let us observe also that for any $(U_0,c_0)$ and $(\delta_U,\delta_c)$ 
the family
$\{\Omega_N(U_0+t \delta_U,c_0+t\delta_c\}_{t\in[0,1]}$ is a convex
one-parameter family and then, according to the Brunn-Minkowski theorem
the function $R_N(t)\equiv V^{1/N}(U_0+t\delta_U,c_0+t \delta_c)$ is
concave. But since in our consideration $N\to\infty$,
 to obtain that this function is strictly concave in some neighbourhood of
the point $(U^*,c^*)$ of maximum of $s_N(U,c)-U$,
 we shall use some corollary from the theorem of Brunn- Minkowski:
\begin{proposition}\label{pro:1}
Consider the convex set ${\cal M}\subset {\bf R^N}$ whose  boundary
consists of a finite number of smooth pieces.
Let the convex one-parameter family $\{{\cal A}(t)\}_{t_1^*\le t\le t_2^*}$
be given by the  intersections of  ${\cal M}$ with the
parallel the hyper-planes
${\cal B}(t)\equiv\{\bJ:\,(\bJ,{\bf e})=tN^{1/2}\}$. Suppose that there
 is some smooth piece   ${\cal D}$ of the boundary of ${\cal M}$, such that
for any $\bJ\in {\cal D}$
 the minimal normal curvature  satisfies the inequality
$N^{1/2}\kappa_{min}(\bJ)>K_0$,  and the Lebesgue measure $S(t)$ of the
intersection ${\cal D}\cap{\cal A}(t)$ satisfies the bound
\begin{equation}
S(t)\ge N^{1/2} V(t)C(t),
\label{p1.0}\end{equation}
where $V(t)$  is the volume of ${\cal A}(t)$.
Then $\frac{d^2}{dt^2}V^{1/N}(t)\le -  K_0C(t)V^{1/N}(t)$.
\end{proposition}

\smallskip

One can see that, if we consider the
sets ${\cal M}, {\cal M}',{\cal A}, {\cal B}(t)\subset {\bf R}^{N+1}$
$$ \begin{array}{c}
{\cal M}\equiv {\cal M}'\cap {\cal A},\,\,\,
{\cal M}'\equiv\{(\bJ,U):\,
NU\ge \Phi_N(\bJ),\,\,\bJ\in\Gamma_N\},\\
{\cal A}\equiv\{(\bJ,U):\,
\delta_U((\bJ,{\bf e})-N^{1/2}c_0)-N^{1/2}\delta_c(U-U_0)=0\},\\
{\cal B}(t)\equiv \{(\bJ,U):\
\delta_c((\bJ,{\bf e})-N^{1/2}c_0)+N^{1/2}\delta_U(U-U_0)=N^{1/2}t\},
\end{array}$$
then $\Omega_N(U_0+t \delta_U,c_0+t\delta_c)={\cal M}\cap {\cal B}(t)$
(without loss of generality we assume that $\delta_c^2+\delta_U^2=1$).
Conditions (\ref{cond1}) and (\ref{cond3})  guarantee that the
minimal normal curvature of ${\cal D}_N'(U)\equiv\{(\bJ,\Phi_N(\bJ)),\,\bJ\in
\Gamma_N\}$ satisfies the inequality
$N^{1/2}\kappa_{min}(\bJ)>\ti K$ for $\bJ\in {\cal D}_N(U)$, if $|U-U^*|<\e$ with
small enough but $N$-independent $\e$. Besides, similarly to (\ref{t1.10})
$$
\frac{\hbox{mes}{\cal D}_N(U,c)}{N^{1/2} V_N(U,c)}
\ge  C_3\frac{d}{dU}s_N(U,c).
$$
 Thus we get that
\begin{equation}
{d\over dU}s_N(U,c)\ge\frac{1}{2}\Rightarrow
{d^2\over dt^2}s_N(U+t\sin\varphi,c+ t\cos\varphi)\bigg|_{t=0}\le-C_4.
\label{t1.11}\end{equation}
\begin{remark}\label{rem:2}
If $\Gamma_N={\bf R}^N$, then conditions of Theorem \ref{thm:1} guarantee
that ${d\over dU}s_N(U,c)\ge const$, when $(U,c)\sim(U^*,c^*)$ and so 
Proposition \ref{pro:1} and (\ref{t1.10}) give us that
\begin{equation}
s_N(U,c)-U-(s_N(U^*,c^*)-U^*)\le -{\ti C_0\over 2}((c-c^*)^2+(U-U^*)^2).
\label{t1.10a}\end{equation}
which implies immediately (\ref{t1.1}). But in the general case, the proof is more 
complicated.
\end{remark} 
Let us introduce the new variables $\rho\equiv((U-U^*)^2+(c-c^*)^2)^{1/2}$,

\no $\varphi\equiv\arcsin\frac{U-U^*}{((U-U^*)^2+(c-c^*)^2)^{1/2}}$ and let
 $\tilde\phi_N(\rho,\varphi)\equiv\phi_N(U,c)\equiv s_N(U^*+U,c^*+c)-U-
 s_N(U^*,c^*)+U^*$. We shall prove  now that
 \begin{equation}
\tilde\phi_N(N^{-1/2},\varphi)\le -\frac{K}{N},
\label{t1.12}\end{equation}
where $K$ does not depend on $\varphi$, $N$.  Consider the set 
$$
\Lambda=\left\{(U,c):  {d\over dU}s_N(U,c)<\frac{1}{2}\right\}.
$$
 One can see easily, 
that if $(U',c')\in\Lambda$, then $(U,c')\in\Lambda$ for any $U>U'$  and 
${d\over dU}\phi_N(U,c')<-\frac{1}{2}$. That is why it is clear, that 
$(U^*,c^*)\not\in\Lambda$ (but it can belong to the boundary $\partial\Lambda$).
Denote 
$$
\varphi^*\equiv\inf_{\varphi\in[-\frac{\pi}{2},\frac{\pi}{2}]}\left\{
\ov r(N^{-1/2}\sin\varphi,N^{-1/2}\cos\varphi)\cap\Lambda\not=\emptyset\right\},
$$ 
where $\ov r(U,c)$ is the set of all points of the form
  $(U^*+tU,c^*+tc)$, $t\in[0,1]$).
Then for any $\varphi<\varphi^*$ we can apply (\ref{t1.11}) to obtain that
\begin{equation}
\tilde\phi_N(N^{-1/2},\varphi)\le -\frac{C_4}{2N}.
\label{t1.13}\end{equation}
Assume that $-\frac{\pi}{4}\le \varphi^* \le\frac{\pi}{4}$. Let us remark that,
 using  (\ref{cond3}), similarly to (\ref{t1.9a}) one can obtain that for all
$(U,c)$: $|U- U^*|\le N^{-1/2}$ and $|c- c^*|\le N^{-1/2}$
 \begin{equation}\begin{array}{c}\ds{
 {d\over dU}s_N(U,c)\le\min|\nabla \Phi_N(\bJ)|^{-1}\frac{S_N(U,c)}{V_N(U,c)}
\le C_5}.
\end{array}\label{t1.14}\end{equation}
Choose $d\equiv\frac{C_4}{4C_5}$. Then for all $\varphi^*\le\varphi\le
\varphi_d\equiv \arctan(\tan\varphi^*+dN^{-1/2})$, using (\ref{t1.13}) and
(\ref{t1.14}), we have got
\begin{equation}\begin{array}{c}\ds{
\tilde\phi_N(N^{-1/2},\varphi)=\phi_N(N^{-1/2}\sin\varphi,N^{-1/2}\cos\varphi)}\\
\ds{
\le \phi_N(N^{-1/2}\sin\varphi-\frac{d}{N},N^{-1/2}\cos\varphi)+
\frac{C_5d}{N}\le
-\frac{C_4}{4N}+O(N^{-3/2})}.
\end{array}\label{t1.15}\end{equation}
For $\frac{\pi}{4}\ge\varphi>\varphi_d$, according to the definition of 
$\varphi^*$ and $\varphi_d$, there exists $\rho_1<1$ such that
$$\begin{array}{c}
(N^{-1/2}\rho_1\sin\varphi-{d\rho_1\over N},\,N^{-1/2}\rho_1\cos\varphi)
\in\Lambda\\
\Rightarrow (N^{-1/2}\rho_1\sin\varphi-{td\rho_1\over N},\,N^{-1/2}\rho_1\cos\varphi)
\in\Lambda
\,\,\, (t\in[0,1]).
\end{array}$$
 Therefore, using that 
$\tilde\phi_N(\rho,\varphi)$ is a concave function of $\rho$, we get
\begin{equation}\begin{array}{c}\ds{
\tilde\phi_N(N^{-1/2},\varphi)\le \rho_1^{-1} 
\tilde\phi_N(N^{-1/2}\rho_1,\varphi)}\\
\ds{
=\rho_1^{-1}\phi_N(N^{-1/2}\rho_1\sin\varphi,N^{-1/2}\rho_1\cos\varphi)}\\
\ds{
\le \rho_1^{-1}\phi_N(N^{-1/2}\rho_1\sin\varphi-
\frac{d\rho_1}{N},N^{-1/2}\rho_1\cos\varphi)-\frac{d}{2N}\le -\frac{d}{2N}}.
\end{array}\label{t1.16}\end{equation}
And finally, if $|\varphi|>\frac{\pi}{4}$, 
denote
$$
{\cal L}_\phi\equiv \ov r(N^{-1/2}\sin\varphi,N^{-1/2}\cos\varphi)\cap\Lambda,
\,\,
l_\phi=N^{1/2}{\hbox{ mes}}\{{\cal L}_\phi\}.
$$
Then, using that for $(U,c)\in {\cal L}_\phi$ 
$$\frac{d}{d\rho}\ti\phi_N
 (N^{-1/2}\rho,\varphi)\le N^{-1/2}\cos\frac{\pi}{4}\frac{d}{dU}
 \phi_N(U,c) <-\frac{1}{2}N^{-1/2}\cos\frac{\pi}{4},
 $$ 
and for  $(U,c)\not\in {\cal L}_\phi$ we can apply (\ref{t1.11}), we have got
\begin{equation}
\tilde\phi_N(N^{-1/2},\varphi)\le -\frac{(1-l_\phi)^2C_4}{2N}
-\frac{l_\phi}{2(2N)^{1/2}}\le -\frac{K}{N}.
\label{t1.17}\end{equation} 
Inequalities (\ref{t1.13})-(\ref{t1.17}) prove (\ref{t1.12}) for 
$|\varphi|<\frac{\pi}{2}$. For the rest of $\varphi$ the proof is the same.

Now let us derive (\ref{t1.1}) (for $p=2$) from (\ref{t1.12}). Choose 
$\rho^*=\frac{4}{K}$ and remark, that since  $\tilde\phi_N(\rho,\varphi)$ 
is a concave function of $\rho$, we  have got that 
 for $\rho>N^{-1/2}\rho^*$ 
$$
\frac{1}{2}\frac{d}{d\rho}\tilde\phi_N(\rho,\varphi)\bigg|_{\rho=N^{-1/2}\rho^*}
\le \frac{d}{d\rho}[\tilde\phi_N(\rho,\varphi)+\frac{2}{N}\log\rho]<
 -\frac{K}{2N^{1/2}}. 
$$ 
Thus, using the Laplace method, one can obtain that
$$
\frac{\int_{\rho>N^{-1/2}\rho^*}d\rho\,
\rho^2e^{N\tilde\phi_N(\rho,\varphi)}}
{\int_{\rho>N^{-1/2}\rho^*}d\rho e^{N\tilde\phi_N(\rho,\varphi)}}\le 
\frac{(\rho^*)^2}{N}
\frac{\frac{d}{d\rho}\tilde\phi_N(\rho,\varphi)}
{\frac{d}{d\rho}[\tilde\phi_N(\rho,\varphi)+
\frac{2}{N}\log\rho]}
\Bigg|_{\rho=N^{-1/2}\rho^*}\le \frac{2(\rho^*)^2}{N}.
$$
So, we have for any $\varphi$
$$\begin{array}{c}\ds{
\int d\rho\,\rho^2e^{N\tilde\phi_N(\rho,\varphi)}\le
\frac{(\rho^*)^2}{N}
 \int_{\rho<N^{-1/2}\rho^*}d\rho\, e^{N\tilde\phi_N(\rho,\varphi)}}\\
 \ds{
 +  \frac{2(\rho^*)^2}{N}
 \int_{\rho>N^{-1/2}\rho^*}d\rho\, e^{N\tilde\phi_N(\rho,\varphi)}\le
2\frac{(\rho^*)^2}{N} \int d\rho 
e^{N\tilde\phi_N(\rho,\varphi)}.}
\end{array}$$
This relation proves (\ref{t1.1}) for $p=2$, because of the inequalities
$$
\la(c-\la c\ra_{(U,c)})^2\ra_{(U,c)}\le \la(c-c^*)^2\ra_{(U,c)}\le
\frac{\int d\phi\int d\rho\,\rho^2e^{N\tilde\phi_N(\rho,\varphi)}}
{\int d\phi\int d\rho e^{N\tilde\phi_N(\rho,\varphi)}}\le\frac{2(\rho^*)^2}{N}.
$$ 
For  other values of $p$ the proof  of (\ref{t1.1}) is similar.
\bigskip

\no{\it Proof of Theorem \ref{thm:2}}

For our consideration below it is convenient to introduce also
the Hamiltonian
\begin{equation}
\ov\H_{N,p}(\bJ,\vx,h,z,\e)\equiv{1\over 2\e}\sum_{\mu=1}^p(N^{-1/2}(\bxm,\bJ)
-x^{(\mu)})^2+h(\bh,\bJ)+
{z\over 2}(\bJ,\bJ).
\label{H(x)}\end{equation}

Evidently
$$
\H_{N,p}(\bJ,k,h,z,\e)=-\log\int_{\xm>k}d\vx\exp\{\ov\H_{N,p}(\bJ,\vx,h,z,\e)\}
+\frac{p}{2}\log (2\pi\e)
$$
and so  $\la \ti F(\bJ)\ra=\la \ti F(\bJ)\ra_{\ov\H_{N,p}}$ for 
any $\ti F(\bJ)$.
Therefore below we denote $\la\dots\ra$ both averaging with respect to
${\cal H}_{N,p}$ and $\ov {\cal H}_{N,p}$.

\begin{lemma}\label{lem:1}
Define the matrix $X_N^{\mu,\nu}=\ds{\frac{1}{N}\sum_{i=1}^N\xmi\xni}$. If
the inequalities
\begin{equation}\begin{array}{c}\ds{
||X_N||\le (\sqrt\alpha+2)^2,\quad
\frac{1}{N}({\bf h},{\bf h})\le 2},
\end{array}\label{matr_X}\end{equation}
are fulfilled, then the Hamiltonian $\H_{N,p}(\bJ,k,h,z,\e)$  satisfies conditions
(\ref{cond1}), (\ref{cond2}), (\ref{cond3}) and (\ref{cond4}) of
Theorem \ref{thm:1} and therefore
\begin{equation}\begin{array}{c}\ds{
{1\over N^2}\sum_{i,j=1}^N
\la\dot J_i\dot J_j\ra\la J_i\ra\la J_j\ra\le
 {C(z,\e)\over N},\quad
{1\over N^2}\sum_{i,j=1}^N
\la\dot J_i\dot J_j\ra^2\le
 {C(z,\e)\over N},}
\end{array}\label{main}\end{equation}
where $\dot J_i\equiv J_i-\la J_i\ra$.

Moreover, choosing $\e_N\equiv N^{-1/2}\log N$ we have got that there exist
$N$-independent $C_1$ and $C_2$, such that
\begin{equation}
\emph{Prob}\left\{\max_i\la\theta(J_i-N^{1/2} \e_N)\ra>e^{-C_1\log^2N}
\right\}\le e^{-C_2\log^2 N}.
\label{bound_J}\end{equation}
\end{lemma}

\begin{remark}\label{rem:3}
 According to the result of \cite{ST1} and to a low of large numbers,
 $P_N$-the probability that inequalities (\ref{matr_X}) are
fulfilled, is more than  $1-e^{-\emph{const} N^{2/3}}$
\end{remark}

\begin{remark}\label{rem:4}
Let us note that since the Hamiltonian (\ref{H_N,p}) under
conditions (\ref{matr_X}) satisfies
(\ref{cond1}), (\ref{cond2}) and (\ref{cond4}), we can choose $R_0$
large enough to have
$$\begin{array}{c}\ds{
\sigma_N^{-1}\int_{\Gamma_N}\theta(|\bJ|-N^{1/2}R_0)e^{-{\cal H}_{N,p}}d\bJ\le
(R_0)^N e^{-NC_1R_0^2}<e^{-NC_3-N}}\\
\Rightarrow
\lla \theta(|\bJ|-N^{1/2}R_0)\rra\le e^{-N},
\end{array}$$
so in all computations below we can use the inequality
$|\bJ|\le N^{1/2}R_0$ with the error
 $O(e^{-N\const})$.
\end{remark}

\begin{remark}\label{rem:5}
Let us note, that   sometimes it is convenient to use  (\ref{main}) in the form
$$\begin{array}{c}\ds{
E\biggl\{\biggl\langle\biggl(N^{-1}\sum_{i}^N\dot J_i^{(1)}\dot J_i^{(2)} \biggr)^2
\biggr\rangle^{(1,2)}\biggr\}\le
 {C(z,\e)\over N},}\\
 \ds{
E\biggl\{\biggl\langle\biggl(N^{-1}\sum_{i}^N\dot J_i\la J_i
\ra\biggr)^2\biggr\rangle
\biggr\}\le
 {C(z,\e)\over N}.}
\end{array}$$
Here and below we  put an upper index
to $J_i$ to show that we take a few replicas of our Hamiltonians and
the upper index indicate the  replica number. We put also
an upper index $\la..\ra^{(1,2)}$ to stress that we consider the Gibbs measure
for two replicas.
The last relations means, in particularly, that
\begin{equation}
{1\over N}\sum\dot J_i^{(1)}\dot J_i^{(2)}\to 0,\,\,\,{1\over N}\sum\dot J_i\la J_i\ra
\to 0,\,\,\, as\,\,\, N\to\infty
\label{main2}\end{equation}
in the Gibbs measure and the probability.
\end{remark}

We start the proof of Theorem \ref{thm:2}
 from the proof of the self-averaging property (\ref{s-a}).
of $f_{N,p}(h,z,\e)$.
Using the idea, proposed in \cite{PS} (see also \cite{ST1}), we write
$$
f_{N,p}(h,z,\e)-E\{f_{N,p}(k,h,z,\e)\}=
{1\over N}\sum_{\mu=0}^p\Delta_\mu,
$$
where
$$
\Delta_\mu\equiv E_\mu\left\{(\log Z_{N,p}(k,h,z,\e))\right\}-
E_{\mu+1}\left\{(\log Z_{N,p}(h,z,\e))\right\},
$$
the symbol $E_\mu\{..\}$ means the averaging with respect to
random vectors $\bx1,...,\bxm$ and

\no $E_0\left\{\log Z_{N,p}(k,h,z,\e)\right\}=
\log Z_{N,p}(h,z,\e)$. Then, in the  usual way,
$$
E\left\{\Delta_\mu\Delta_\nu\right\}=0 \,\, (\mu\not=\nu),
$$
and therefore
\begin{equation}
E\left\{(f_{N,p}(h,z,\e)-E\{f_{N,p}(k,h,z,\e)\})^2\right\}=
{1\over N^2}\sum_{\mu=0}^pE\{\Delta_\mu^2\}.
\label{s-a.2}\end{equation}
But

\begin{equation}\begin{array}{c}\ds{
E\{\Delta_{\mu-1}^2\}\le E\{(E_{\mu-1}\{(\log Z_{N,p}(k,h,z,\e))\} }\\
\ds{
\-E_{\mu-1}\left\{(\log Z_{N,p-1}^{(\mu)}(k,h,z,\e)))^2\right\}\le
E\{(\Delta_\mu')^2\},}
\end{array}\label{s-a.3}\end{equation}
where
$$
\Delta_\mu'\equiv\log Z_{N,p}(k,h,z,\e)-
\log Z_{N,p-1}^{(\mu)}(k,h,z,\e),
 $$
 with $ Z_{N,p-1}^{(\mu)}(k,h,z,\e)$ being the partition function for
 the  Hamiltonian (\ref{H_N,p}), where in the r.h.s. we take the sum with
 respect to all upper indexes except $\mu$. Denoting by 
 $\la...\ra_{p-1}^{(\mu)}$ the correspondent Gibbs averaging and
 integrating with respect to $\vx$, we get:
 \begin{equation}\begin{array}{c}\ds{
\Delta_\mu'=\sqrt\e\log\lla  \fH\left(\frac{
k-(\bxm,\bJ) N^{-1/2}}{\sqrt\e}\right)\rra_{p-1}^{(\mu)}.}
\end{array}\label{s-a.4}\end{equation}
But evidently
\begin{equation}\begin{array}{c}
0\ge\log\lla \fH\left(\e^{-1/2}(
k-(\bxm,\bJ) N^{-1/2})\right) \rra_{p-1}^{(\mu)}\\
\ds{
\ge\lla\log \fH\left(\e^{-1/2}(
k-(\bxm,\bJ) N^{-1/2})\right)\rra_{p-1}^{(\mu)} }\\
\ds{
\ge-\const\lla (N\e)^{-1}(\bxm,\bJ)^2 \rra_{p-1}^{(\mu)}+\const}
\end{array}\label{s-a.5}\end{equation}
Thus,
$$
E\{(\Delta_\mu')^2\}\le\const E\left\{\lla (N\e)^{-1} (\bxm,\bJ)^2
\rra_{p-1}^{(\mu)}
\lla (N\e)^{-1}(\bxm,\bJ)^2\rra_{p-1}^{(\mu)}\right\}.
$$

But since $\la...\ra_{p-1}^{(\mu)}$ does not depend on $\bxm$ we can average with respect to
$\bxm$ inside $\la...\ra_{p-1}^{(\mu)}$. Hence, we obtain
\begin{equation}
E\{(\Delta'_\mu)^2\}\le \const\e^{-2}E\left\{\lla N^{-1}(\bJ,\bJ)
\rra_{p-1}^{(\mu)}
\lla N^{-1}(\bJ,\bJ)\rra_{p-1}^{(\mu)}\right\}\le\const.
\label{s-a.6}\end{equation}
Inequalities (\ref{s-a.2})-(\ref{s-a.5})
prove (\ref{s-a}).

\medskip

Define the order parameters of our problem
\begin{equation}
 R_{N,p}\equiv{1\over N}\sum_{i=1}^N \la J_i^2\ra,\quad
  q_{N,p}\equiv{1\over N}\sum_{i=1}^N \la J_i\ra^2
\label{def_R,q}\end{equation}
To prove the self-averaging properties of $R_{N,p}$ and $q_{N,p}$
we use the following general lemma:

\begin{lemma} \label{lem:2}
Consider the sequence of convex random functions $\{f_n(t)\}_{n=1}^\infty$
($f_n''(t)\ge 0$) in the interval $(a,b)$.
If functions $f_n$  are self-averaging 
($E\{(f_n(t)-E\{f_n(t)\})^2\}\to 0$, as $n\to\infty$ uniformly in $t$)
and bounded ($|E\{f_n(t)\}|\le C$ uniformly in $n$, $t\in (a,b)$),
then for almost all  $t$
\begin{equation}  \begin{array}{c}\ds{
\lim_{n\to\infty} E \{ [f_{n}'(t) - E\{f'_n(t)\} ]^2 \} = 0 ,}
\end{array}
\label{II.3.3}\end{equation}
i.e. the derivatives $f_n'(t)$  are also self-averaging ones  for almost
all $t$.

In addition, if we consider  another sequence of  convex functions
 $\{g_n(t)\}_{n=1}^\infty$ ($g_n''\ge 0$)
which  are also self-averaging 
($E\{(g_n(t)-E\{g_n(t)\})^2\}\to 0$, as $n\to\infty$ uniformly in $t$),
 and $|E\{f_n(t)\}-E\{g_n(t)\}|\to 0$, as $n\to\infty$,
uniformly in $t$, then for all $t$,  which satisfy (\ref{II.3.3})
\begin{equation}
\lim_{n\to\infty}|E\{f_n'(t)\}-E\{g_n'(t)\}|=0,\qquad
\lim_{n \to \infty} E \{ [ g_{n}'(t) -E\{ g'(t)\}  ]^2 \} = 0 .
\label{II.3.3b}\end{equation}
\end{lemma}
For the proof of this lemma see \cite{PST2}. On the basis of Lemma
\ref{lem:2}, in Sec.4 we prove

\begin{proposition}\label{pro:2}
Denote
$R_{N,p-1}$, $q_{N,p-1}$ the analogs  of $R_{N,p}$, $q_{N,p}$
(see definition (\ref{def_R,q})) for $H_{N,p-1}$. Then for any convergent
subsequence $E\{f_{N_m,p_m}(k,h,z,\e)\}$ for almost all
$z$ and $h$ $ R_{N_m,p_m}$, $q_{N_m,p_m}$ we have got
\begin{equation}\begin{array}{c}\ds{
E\{( R_{N_m,p_m}-\ov R_{N_m,p_m})^2\},\, E\{( q_{N_m,p_m}-\ov q_{N_m,p_m})^2\}
\to 0,}\\
\ds{
|\ov R_{N_m,p_m}-\ov R_{N_m,p_m-1}|,\, |\ov q_{N_m,p_m}-\ov q_{N_m,p_m-1}|
\to 0\,\,\ as\,\,\, k\to\infty,}
\end{array}\label{p2.1}\end{equation}
where
\begin{equation}
\ov R_{N,p}=E\{R_{N,p}\}, \quad \ov q_{N,p}=E\{ q_{N,p}\}
\label{ov-R}\end{equation}
and
\begin{equation}
E\biggl\{\biggl\langle\biggl( N_m^{-1} \sum_{i=1}^{N_m} J_i^2-\ov R_{N_m,p_m}
\biggr)^2\biggr\rangle
\biggr\} \to 0,\,\,  as\,\, N_m\to\infty.
\label{s-a.R}\end{equation}
\end{proposition}
 Our strategy now is to choose an arbitrary convergent subsequence
$f_{N_m,p_m}(k,h,z,\e)$, by  applying to it the above proposition, to show that its limit
for all $h,z$ coincides with the r.h.s. of (\ref{t2.1}). Then this will mean
that there exists the limit $f_{N,p}(h,z,\e)$ as $N,p\to\infty$,
${p\over N}\to\a$. But in order to simplify formulae below we shall
omit the subindex $m$ for $N$ and $p$.

\medskip

Now we formulate the main technical point of the proof of Theorem \ref{thm:2}.

\begin{lemma}\label{lem:3}
Consider  $H_{N,p-1}$ and denote by $\la\dots\ra_{p-1}$ the respective Gibbs
averages. For any $\e_1>0$ and $0\le k_1\le 2k$ define
\begin{equation}
\phi_N(\e_1,k_1)\equiv \e_1^{1/2}\lla \emph{H}\biggl({k_1-N^{-1/2}(\bxp,\la\bJ\ra_{p-1})\over
\sqrt{\e_1}}\biggr) \rra_{p-1},
\label{l3.1}\end{equation}
\begin{equation}
\phi_{0,N}(\e_1,k_1)\equiv
\e_1^{1/2}\emph{H}\biggl({k_1-N^{-1/2}(\bxp,\la\bJ\ra_{p-1})\over
\sqrt{U_{N,p-1}(\e_1)}}\biggr),
\label{l3.2}\end{equation}
where
$\quad U_{N,p-1}(\e_1)\equiv \ov R_{N,p-1}-\ov q_{N,p-1}+\e_1$.
Then,
\begin{equation}\begin{array}{c}
E\left\{\left(\phi_{N}(\e_1,k_1)-\phi_{0,N}(\e_1,k_1)\right)^2\right\}
\to 0,\\
E\left\{\left(\log\phi_{N}(\e_1,k_1)-\log\phi_{0,N}(\e_1,k_1)\right)^2
\right\}\to 0,\\
E\biggl\{\biggl({d\over d\e_1}\log\phi_{N}(\e_1,k_1)-{d\over d\e_1}
\log\phi_{0,N}(\e_1,k_1)\biggr)^2\biggr\}\to 0,\\
E\biggl\{\biggl({d\over dk_1}\log\phi_{N}(\e_1,k_1)-{d\over dk_1}
\log\phi_{0,N}(\e_1,k_1)\biggr)^2\biggr\}\to 0,
\end{array}\label{l3.3}\end{equation}
and $N^{-1/2}(\bxp,\la\bJ\ra_{p-1})$ converges in distribution
to $\sqrt{\ov q_{N,p}}u$, where  $u$ is a Gaussian random variable with zero
mean and variance 1.

Besides, if we denote
 \begin{equation}\begin{array}{c}\ds{
t^{(\mu)}\equiv N^{-1/2}(\bxm,J)-x^\mu,\quad \dot\tm\equiv\tm-\la\tm\ra}\\
\ds{
\ti U_{N}\equiv{1\over \e^2 N}\sum_{\mu=1}^p
\la (\tm)^2\ra,\quad
\ti q_{N}\equiv {1\over \e^2 N}\sum_{\mu=1}^p\la\tm\ra^2,}
\end{array}\label{ti-q}\end{equation}
then $\ti U_N$ and $\ti q_N$ are self-averaging quantities  and
for $\mu\not=\nu$
 \begin{equation}\begin{array}{c}\ds{
 E\left\{\la\dot\tm\dot\tn\ra^2\right\}\to 0,\,\,\,
E\left\{\la((\tm)^2-\la(\tm)^2\ra) ((\tm)^2-\la(\tm)^2\ra)\ra^2\right\}\to 0,}\\
 \ds{
E\left\{\la(\tm)^4\ra\right\}\le\const,\,\,\,
E\left\{\la(\tm)^4 (\tn)^4\ra\right\}\le\emph{const}.}
\end{array}\label{l3.4}\end{equation}
\end{lemma}

Now we are ready to derive the equations for $\ov q_{N,p}$ and $\ov R_{N,p}$.
From the symmetry of the Hamiltonian (\ref{H(x)}) it is evident that
$\ov q_{N,p}=E\{\la J_1\ra^2\}$ and $\ov R_{N,p}=E\{\la J_1^2\ra\}$.
The integration with respect $J_1$ is Gaussian. So, if we denote
$$
\tm_1\equiv\tm-N^{-1/2}\xmo J_1,
$$
 we get
$$
\la J_1\ra=-(z+\a_N/\e)^{-1}\biggl({1\over \e N^{1/2}}\sum_{\mu=1}^p\xmo \la\tm_1\ra
+hh_1\biggr).
$$
Hence,
\begin{equation}\begin{array}{c}\ds{
(z+\a_N/\e)^2E\left\{\la J_1\ra^2\right\}=
{1\over \e^2N}E\biggl\{\sum_{\mu,\nu=1}^p\xmo \xno\la\tm_1\ra\la\tn_1\ra
\biggr\}}\\
\ds{
+h^2+{2h\over \eN^{1/2}}E\biggl\{\sum_{\mu=1}^p h_1\xmo \la\tm_1\ra\biggr\}
+o(1)},
\end{array}\label{exp_q}\end{equation}
and similarly
\begin{equation}\begin{array}{c}\ds{
(z+\a_N/\e)^2E\left\{\la J_1^2\ra\right\}= (z+\a_N/\e)+{1\over \e^2N}
\sum_{\mu,\nu=1}^p E\left\{\xmo \xno \la\tm_1 \tn_1\ra\right\}}\\
\ds{
+h^2+{2h\over \eN^{1/2}}\sum_{\mu=1}^pE\left\{ h_1\xmo \la\tm_1\ra\right\}
+o(1).}
\end{array}\label{exp_R}\end{equation}

 Now to calculate the r.h.s. in (\ref{exp_q}) and (\ref{exp_R}) we use the formula of
 "integration by parts" which is valid for any function $f$ with bounded third derivative
\begin{equation}\begin{array}{c}\ds{
E\left\{\xmo  f\left(\xmo N^{-1/2}\right)\right\}}\\
\ds{ 
= {1\over N^{1/2}}E\left\{f'\left(\xmo N^{-1/2}\right)\right\}
+{1\over N^{3/2}}E\left\{f'''\left(\zeta(\xmo)\xmo N^{-1/2}
\right)\right\},}
\end{array}\label{ibp}\end{equation}
where $|\zeta(\xmo)|\le 1$.
Thus, using this formula and the second line of (\ref{l3.4}), we get:
\begin{equation}\begin{array}{c}\ds{
(z+\a_N/\e)^2\ov q_{N,p}=\ti q_N}\\
\ds{
+{1\over N^2\e^4}\sum_{\mu\not=\nu}E\left\{
\lla\dot\tm_1(\tm_1J_1-\la \tm_1J_1\ra)\rra\lla\dot\tn_1(\tn_1J_1-
\la \tn_1J_1\ra)\rra\right\}}\\
\ds{
+{2\over N^2\e^4}\sum_{\mu\not=\nu}E\left\{
\lla\dot\tm_1(\tm_1J_1-\la \tm_1J_1\ra)(\tn_1J_1-\la \tn_1J_1\ra)\rra
\la\tn_1\ra\right\} }\\
\ds{
+{1\over N^2\e^4}\sum_{\mu\not=\nu}E\left\{
\lla\dot\tm_1(\tn_1J_1-\la \tn_1J_1\ra)\rra\lla\dot\tn_1(\tm_1J_1-
\la \tm_1J_1\ra)\rra\right\}}\\
\ds{
+h^2+{2h^2\over\e^2N}\sum_{\mu}E\left\{ \lla\dot\tm_1(\tm_1J_1-
\la \tm_1J_1\ra)\dot J_1\rra\right\}+o(1).}
\end{array}\label{q.1}\end{equation}
 Substituting $\tm_1$ by $\tm$ and using the symmetry of the Hamiltonian
 with respect to $J_i$, we obtain e.g. for the first sum in (\ref{q.1}):
 $$\begin{array}{c}\ds{
 {1\over N^2}\sum_{\mu\not=\nu}E\left\{
\lla\dot\tm_1(\tm_1J_1-\la \tm_1J_1\ra)\rra\lla\dot\tn_1(\tn_1 J_1
-\la \tn_1J_1\ra)\rra\right\}}\\
\ds{
={1\over N^3}\sum_{i=1}^N\sum_{\mu,\nu=1}^pE\left\{
\lla\dot\tm(\tm (\dot J_i+\la J_i\ra)-\la \tm (\dot J_i+\la J_i\ra)\ra\rra
\right.}\\
\ds{\left.
\cdot\lla \dot\tn (\dot J_i+\la J_i\ra)-\la \tn (\dot J_i+\la J_i\ra)\ra
\rra\right\}+o(1)}\\
\ds{
={1\over N^3}\sum_{i=1}^N\sum_{\mu,\nu=1}^p E\left\{\la J_i\ra^2
\la(\dot \tm)^2\ra \la(\dot \tn)^2\ra\right\}+o(1)=
\ov q_{N,p}(\ti U_N-\ti q_N)^2+o(1).}
 \end{array}$$
 Here we have used the relation (\ref{main2}), which allows us to get rid from
 the terms containing $\dot J_i$ and the self-averaging properties of
 $\ov q_{N,p}$, $\ti U_N$ and $\ti q_N$. Transforming in a similar
 way the other sums in the r.h.s. of (\ref{q.1}) and using also relations
 (\ref{l3.4}) to get rid from the terms, containing $\la\dot \tm\dot\tn\ra$,
we get finally:
\begin{equation}\begin{array}{c}\ds{
(z+\a_N/\e)^2\ov q_{N,p}=\ti q_N+2(\ov R_{N,p}-\ov q_{N,p})\ti q_N(\ti U_N-\ti q_N)}\\
\ds{
+\ov q_{N,p}(\ti U_N-\ti q_N)^2+h^2(1+2(\ti U_N-\ti q_N)(\ov R_{N,p}-\ov q_{N,p}))
+o(1).}
\end{array}\label{q.2}\end{equation}

Similarly we obtain
\begin{equation}\begin{array}{c}\ds{
(z+\a_N/\e)^2\ov R_{N,p}= (z+\a_N/\e)+\ti U_N+\ov R_{N,p}(\ti U_N^2-\ti q_N^2)}\\
\ds{
-2\ov q_{N,p}\ti q_N(\ti U_N-\ti q_N)+h^2(1+2(\ti U_N-\ti q_N)
(\ov R_{N,p}-\ov q_{N,p}))+o(1).}
\end{array}\label{R.1}\end{equation}
Considering (\ref{q.2}) and (\ref{R.1}) as a system of equation with respect
to $ \ov R_{N,p}$ and $ \ov q_{N,p}$, we get

\begin{equation}\begin{array}{c}\ds{
\ov q_{N,p}={\ti q_N+h^2\over (z+\D)^2}+o(1),\quad
\ov R_{N,p}-\ov q_{N,p}={1\over z+\D}+o(1),}
\end{array}\label{R,q.2}\end{equation}
where we denote for simplicity
\begin{equation}\begin{array}{c}\ds{
\D\equiv {\a\over \e}-\ti U_{N}+\ti q_{N}.}
\end{array}\label{D}\end{equation}

Now  we should find the expressions for $\ti q_{N}$ and $\ti U_{N}$.

From the symmetry of the Hamiltonian (\ref{H_N,p}) it is evident that
\begin{equation}\begin{array}{c}\ds{
\ti q_{N}= \a_NE\left\{{1\over \e^2 }\lla N^{-1/2}(\bxp,\bJ)-x^{(p)}\rra^2
\right\} }\\
\ds{
\left.=\a_NE\left\{\left[{d\over dk_1}\log
\int_{x>0}dx\lla\exp\{-{1\over 2\e_1}(N^{-1/2}(\bxp,\bJ)-x-k_1)^2\}\rra_{p-1}
\right]^2\right\}\right|_{k_1=k}}\\
\ds{
\left.=\a_NE\left\{\left[{d\over dk_1}\log\phi_{N}(k_1,\e_1)\right]^2
\right\}\right|_{k_1=k} .}
\end{array}\label{ti-q.1}\end{equation}
Therefore, using Lemma \ref{lem:3},
we derive:
\begin{equation}\begin{array}{c}\ds{
\ti q_{N}= \a_NE\left\{\left[{d\over dk_1}
\log \fH\left({\sqrt{\ov q_{N,p}}u+k_1\over\sqrt {U_{N,p}}}\right)\right]^2
\right\}={\a_N\over U_{N,p}}E\left\{\fA^2\left({\sqrt{\ov q_{N,p}}u+k_1\over
\sqrt {U_{N,p}}}\right)\right\}.}
\end{array}\label{ti-q.2}\end{equation}
Here and below we denote
\begin{equation}
\fA(x)\equiv -{d\over dx}\log \fH(x)={e^{-x^2/2}\over\sqrt{2\pi}\fH(x)},
\label{A}\end{equation}
where the function $\fH(x)$ is defined by (\ref{H}).
Similarly
\begin{equation}\begin{array}{c}\ds{
\ti U_{N}= \a_NE\left\{{1\over \e^2 }\la(
 N^{-1/2}(\bxp,\bJ)-x^{(p)})^2\ra\right\} }\\
\ds{
\left.=2\a_NE\left\{{d\over d\e_1}\log
\int_{x>0}dx\la\exp\{-{1\over 2\e_1}(N^{-1/2}(\bxp,\bJ)-x-k_1)^2\}\ra_{p-1}
\right\}\right|_{\e_1=\e}}\\
\ds{
\left.=2\a_NE\left\{{d\over d\e_1}\log\phi_p(k_1,\e_1)
\right\}\right|_{\e_1=\e} .}
\end{array}\label{ti-U.1}\end{equation}
Now, using Lemma \ref{lem:3} and Lemma \ref{lem:1},
we derive:
\begin{equation}\begin{array}{c}\ds{
\left.\ti U_{N}=2\a_NE\left\{{d\over d\e_1}\log \e_1^{-1/2}
\fH\left({\sqrt{\ov q_{N,p}}u+k_1\over
 \sqrt {U_{N,p}}}\right)\right\}\right|_{\e_1=\e} }\\
 \ds{
 ={\a_N\over \e}+{\a_N\over U_{N,p}^{3/2}}E\left\{(k+\sqrt{\ov q_{N,p}}u)
\fA\left({ \sqrt{\ov q_{N,p}}u+k_1\over\sqrt {U_{N,p}}}\right)\right\}.}
\end{array}\label{ti-U.2}\end{equation}

Thus, from (\ref{q.2}), (\ref{R.1}), (\ref{ti-q.2}) and (\ref{ti-U.2}) we obtain
the system of equations for $\ov R_{N,p}$ and $\ov q_{N,p}$
\begin{equation}\begin{array}{c}\ds{
\ov q_{N,p}\equiv (\ov R_{N,p}-\ov q_{N,p})^2\bigg[
{\alpha\over U_{N,p}}E\left\{\fA^2\left({\sqrt{\ov q_{N,p}}u+k\over
\sqrt {U_{N,p}}}\right)\right\}+h^2\bigg]+\ti\e_N}\\
\ds{
{\alpha \over U_{N,p}^{3/2}}E\left\{(\sqrt{\ov q_{N,p}}u+k)
\fA\left({\sqrt{\ov q_{N,p}}u+k\over \sqrt {U_{N,p}}}\right)\right\}}\\
\ds{
= z+{\ov q_{N,p}\over(\ov R_{N,p}-\ov q_{N,p})^2}-
{1\over\ov R_{N,p}-\ov q_{N,p}} -h^2+\ti\e_N',}
\end{array}\label{q,R}\end{equation}

where $\ti\e_N, \ti\e_N'\to 0$, as $N,p\to\infty$, $\a_N\to\a$.
\begin{proposition}\label{pro:3}
For any $\a<2$ there exists $\e^*(\alpha,k)$ such that for any
$\e\le\e^*$ and $z<\e^{-1/3}$ the solution of the system (\ref{q,R}) tends
as $\ti\e_N, \ti\e_N'\to 0$ to  $(R^*,q^*)$ which gives the unique point of
$\max_R\min_q$ in the r.h.s. of (\ref{t2.1}).
\end{proposition}

On the basis of this proposition we conclude that for almost all $z,h$
 there exist the limits
$$ \begin{array}{c}\ds{
\lim_{m\to\infty}E\left\{ {d\over dz}f_{N_m,p_m}(k,h,z,\e)\right\}
=R^*(\a,k,h,z,\e),\quad}\\
\ds{
\lim_{m\to\infty}E\left\{ {d\over dh}f_{N_m,p_m}(k,h,z,\e)\right\}
=h(R^*(\a,k,h,z,\e)-q^*(\a,k,h,z,\e)).}
\end{array}$$
But since the r.h.s. here are continuous functions of $z,h$ we derive that
for any convergent subsequence $f_{N_m,p_m}(k,h,z,\e)$ the above limits
exist for all $z,h$. Besides, choosing subsequence $f_{N_m',p_m'}(k,h,z,\e)$
which converges for any rational $\alpha$, we obtain
that for any $N_m',p_m'$ such that $\a_{m}={p_m'\over N_m'}\to\alpha_1$
($\a_1$ is a rational number) and $p_m''$ such that 
$\a'_{m}={p_m''\over N_m'}\to 0$
$$\begin{array}{c}\ds{
E\left\{f_{N_m',p_m'}
(\a_{k},k,h,z,\e)\} -E\{f_{N_m',p_m''}(\a'_{k},k,h,z,\e)\right\}=}\\
\ds{
{1\over N_m'}\sum_{i=0}^{p_m'-p_m''} E\left\{\log Z_{N_m',p_m''-i}(k,h,z,\e)-
\log Z_{N_m',p_m'-i-1}(k,h,z,\e)\right\}}\\
\ds{
\to {1\over N_m'}\sum_{i=0}^{p_m'-p_m''} E\left\{\log
\fH\left({\sqrt{\ov q_{N_m',p_m'-i}}u+k\over
\sqrt {U_{N_m',p_m'-i}}}\right)\right\}}\\
\ds{
\to\int_{0}^{\a_1}E\left\{\log \fH\left({\sqrt{\ov q^*(\a)}u+k\over
\sqrt {R^*(\a)+\e-q^*(\a)}}\right)\right\}d\a}.
\end{array}$$

Thus, for all rational $\a$ there exists
$$
\lim_{m\to\infty}E\left\{f_{N_m,p_m}(k,h,z,\e)\right\}= F(\a,k,h,z,\e),
$$
where  $ F(\a,k,h,z,\e)$ is defined by (\ref{t2.1}). But since
the free energy is obviously monotonically decreasing in $\a$, we  obtain, that
for any convergent subsequence the limit of the free energy coincides
with the r.h.s. of (\ref{t2.1}). Hence, as it was already  mentioned
after Proposition \ref{pro:2}, there exist a limit
which coincides with the r.h.s. of (\ref{t2.1}).
Theorem \ref{thm:2} is proven.

\bigskip

\no{\it Proof of Theorem \ref{thm:3}.}
For any $z>0$ let us take $h$ small enough and consider
$$
\Theta_{N,p}(k,h,z)\equiv\sigma_N^{-1}\int_{\Omega_N} d\bJ\exp\{-{z\over 2}(\bJ,\bJ)-
h({\bf h},\bJ)\},
$$
where
$$
\Omega_{N,p}\equiv\left\{\bJ:\,\, N^{-1/2}(\bxn,\bJ)\ge k,\,\,
(\nu=1,\dots,p)\right\}.
$$
To obtain the self-averaging of $N^{-1}\log_{(MN)}\Theta(k,h,z)$ and
the expression for $E\{N^{-1}\log_{(MN)}\Theta(k,h,z)\}$
we define also the interpolating Hamiltonians, corresponding partition
functions and free energies:
\begin{equation}
{\cal H}^{(\mu)}_{N,p}(\bJ,k,h,z,\e)\equiv-\sum_{\nu=\mu+1}^{p}
\log \fH\left(\frac{k-N^{-1/2}(\bxn,\bJ)}{\sqrt\e}\right)+{z\over 2}(\bJ,\bJ)
+h({\bf h},\bJ),
\label{t3.2}\end{equation}
\begin{equation}\begin{array}{c}
 Z_{N,p}^{(\mu)}(k,h,z,\e)\equiv\sigma_N^{-1}\int_{\Omega_{N,p}^{(\mu)}} d\bJ
\exp\{-{\cal H}^{(\mu)}_{N,p}(\bJ,k,h,z,\e)\},\\
 f_{N,p}^{(\mu)}(k,h,z,\e,M)\equiv{1\over N}\log_{(MN)} Z_{N,p}(k,h,z,\e),
\end{array}\label{t3.4}\end{equation}
where
$$
\Omega_{N,p}^{(\mu)}\equiv\left\{\bJ:\,\,
N^{-1/2}(\mbox{\boldmath$\xi^{(\mu')}$},\bJ)\ge k,\,\,
(\mu'=1,\dots,\mu)\right\}.
$$
According to Theorem \ref{thm:2}, for large enough $M$ with
probability more than $(1-O(N^{-1}))$
$$
 f_{N,p}^{(0)}(k,h,z,\e,M)= f_{N,p}(k,h,z,\e),\quad
 f_{N,p}^{(p)}(k,h,z,\e)={1\over N}\log_{(MN)}\Theta(k,h,z),
$$
where $f_{N,p}(k,h,z,\e)$ is defined by (\ref{f}). Hence,
\begin{equation}\begin{array}{c}\ds{
 f_{N,p}(k,h,z,\e,M)-\frac{1}{N}\log_{(MN)}\Theta_{N,p}(k,h,z)=
{1\over N}\sum_{\mu=1}^{p}\ti\Delta^{(\mu)},}\\
\ds{
\ti\Delta^{(\mu)}\equiv\log_{(MN)}Z^{(\mu-1)}_{N,p}-\log_{(MN)}Z^{(\mu)}_{N,p}.}
\end{array}\label{t3.5}\end{equation}

Below in the proof of Theorem \ref{thm:3} we denote by
$x^{(\mu)}\equiv N^{-1/2}(\bxm,\bJ)$, by
the symbol $\la\dots\ra_{\mu}$ the Gibbs averaging corresponding to the
Hamiltonian  ${\cal H}^{(\mu)}_{N,p}$ in the domain
$\Omega_{N,p}^{(\mu-1)}$ and by $Z^{(\mu,\mu)}_{N,p}$ the correspondent
partition function.Denote also
$$
T_\mu(x)\equiv\lla\theta(x^{(\mu)}-x)\rra_{\mu},\quad
X_\mu\equiv\lla x^{(\mu)}\rra_\mu.
$$

To proceed further, we use the following lemma:
\begin{lemma}\label{lem:4}
If the inequalities (\ref{matr_X}) are fulfilled and
there exists $N,\mu,\e$-independent $D$ such that
\begin{equation}
\frac{1}{N}\lla(\dot{\bJ},\dot{\bJ})\rra_\mu\ge D^2,
\label{cond_D}\end{equation}
then  there exist $N,\mu,\e$-independent $K_1,C_1^*,C_2^*,C_3^*$, such that
for $|X_\mu|\le\log N$
\begin{equation}\begin{array}{c}
T_\mu(k+2\e^{1/4})\ge C_1^* e^{-C_2^*X_\mu^2},\\
T_\mu(k-2\e^{1/4})-T_\mu(k+2\e^{1/4})\le \e^{1/4}C_3^*
\end{array}\label{l4}\end{equation}
with probability $P_N^{(\mu)}\ge(1- K_1N^{-3/2})$.
\end{lemma}

\begin{remark}\label{rem:6}
Similarly to Remark \ref{rem:4} one can conclude  that, if $Z_{N,p}^{(\mu,\mu)}>e^{-MN}$, 
then there exists $\e,N,\mu$-independent $R_0$, such that we can use the inequality
$|\bJ|\le N^{1/2}R_0$ with the error  $O(e^{-N\const})$.
\end{remark}
\begin{remark}\label{rem:7}
Denote $\tilde D_\mu^2$ the l.h.s. of (\ref{cond_D}). Then
$$\begin{array}{c}
4\tilde D_\mu^2\la\theta(|\dot{\bJ}|-2\tilde D_\mu N^{1/2})\ra_\mu\le
N^{-1}\lla(\dot{\bJ},\dot{\bJ})\rra_\mu=\tilde D_\mu^2\\
\Rightarrow
\la\theta(|\dot{\bJ}|-2\tilde D_\mu N^{1/2})\ra_\mu\le {1\over 4}\\
\Rightarrow
Z_N^{(\mu,\mu)}\le {4\over 3}\sigma_N^{-1}\int_{|\dot{\bJ}|<2\tilde D_\mu N^{1/2}}
\exp\{-{z\over 2}(\bJ,\bJ)-h({\bf h},\bJ)\}\le
\frac{4}{3}(2\ti D_\mu)^Ne^{2hNR_0} 
\end{array}$$
 Thus, the inequality  $Z_{N,p}^{(\mu,\mu)}>e^{-MN}$
implies that $\tilde D_\mu\ge{1\over 2}\exp\{-M-2hR_0\}\equiv D^2$.
\end{remark}

Let us prove  self-averaging property of
$f_{N,p}^{(p)}(k,h,z,\e,M)$, using  Lemma \ref{lem:4}. Similarly to (\ref{s-a.2}) we write
$$
f_{N,p}^{(p)}(k,h,z,\e,M)-E\{f_{N,p}^{(p)}(k,h,z,\e,M)\}=
{1\over N}\sum_{\nu=0}^{p-1}\Delta_\nu,
$$
where
$$
\Delta_\nu\equiv E_\nu\{f_{N,p}^{(p)}(k,h,z,\e,M)\}-
E_{\nu+1}\{(f_{N,p}^{(p)}(k,h,z,\e,M)\},
$$
Then $E\{\Delta_\nu\Delta_{\nu'}\}=0$,  $(\nu\not=\nu')$ and therefore
\begin{equation}
E\{( f_{N,p}^{(p)}(k,h,z,\e,M)-E\{f_{N,p}^{(p)}(k,h,z,\e,M)\})^2\}=
{1\over N^2}\sum_{\nu=0}^{p-1}E\{\Delta_\nu^2\},
\label{ti-sa}\end{equation}
where similarly to (\ref{s-a.3})
\begin{equation}
E\{\Delta_\nu^2\}\le E\{\ov \Delta_\nu^2\},
\label{ti-sa.1}\end{equation}
with
$$
\ov \Delta_\nu\equiv
 \log_{(MN)}Z_{N,p}^{(p)}-\log_{(MN)}Z_{N,p}^{(p,\nu+1)},
$$
where $Z_{N,p}^{(p,\nu)}$ is the partition function,
corresponding to the Hamiltonian
${\cal H}^{(p)}_{N,p}$ in the domain
$\Omega_{N,p}^{(p,\nu)}$ which differs from $\Omega_{N,p}^{(p)}$
 by the absence of the inequality
for $\mu'=\nu$. Therefore for $\nu\le p-1$
\begin{equation}\begin{array}{c}
E\{|\ov \Delta_\nu|^2\}=E\{|\ov \Delta_{p-1}|^2\}\\
=E\{ \theta(Z_{N,p}^{(p,p)}-e^{-MN}) |\log_{(MN)}Z_{N,p}^{(p)}-
\log_{(MN)}Z_{N,p}^{(p,p)}|^2\}\\
+E\{ \theta(e^{-MN}-Z_{N,p}^{(p,p)}) |\log_{(MN)}Z_{N,p}^{(p)}-
\log_{(MN)}Z_{N,p}^{(p,p)}|^2\}.
\end{array}\label{ti-sa.1a}\end{equation}
But the second term in the r.h.s. is zero, because $Z_{N,p}^{(p)}
\le Z_{N,p}^{(p,p)}$
and thus $Z_{N,p}^{(p,p)}\le e^{-MN}$ implies $Z_{N,p}^{(p)}\le e^{-MN}$,
and so $\log_{(MN)}Z_{N,p}^{(p)}=
\log_{(MN)}Z_{N,p}^{(p,p)}=-MN$.
 Then, denoting by $\chi_\mu$ the indicator function of the set, where
 $Z^{(\mu,\mu)}>e^{-MN}$, and the inequalities (\ref{l4})  are fulfilled, 
on the basis of Lemma \ref{lem:4},  we obtain that 
\begin{equation}\begin{array}{c}
E\{\ov \Delta_\nu^2\}=E\{\theta(Z_{N,p}^{(p,p)}-e^{-MN}) \log^2_{(M)}\lla
\theta(x^{(p)}-k)\rra_{p}\}\\
\le(MN)^2[E\{\theta(Z_{N,p}^{(p,p)}-e^{-MN}) \theta(|X_p|-\log N)\}\\
+E\{\theta(Z_{N,p}^{(p,p)}-e^{-MN})(1-\chi_p) \theta(\log N-|X_p|)\}]
\\
+E\left\{\theta(Z_{N,p}^{(p,p)}-e^{-MN})\chi_p 
\theta(\log N-|X_p|)\log^2\exp\{-C_1^*X_\mu^2\}
\right\}\\
\le (MN)^2[e^{-\log^2N/2R_0^2})+K_1N^{-3/2}]+2(R_0^2C_1^*)^2\le
2M^2K_1N^{1/2}.
\end{array}\label{ti-sa.2}\end{equation}
Here we have used that, according to the definition of the function
$\log_{(MN)}$ (see (\ref{log_M}), $|\log_{(MN)}\lla \theta(x^{(p)}-k)\rra_{p}|
\le MN$. Besides, we used  the standard Chebyshev
inequality, according to which
\begin{equation}
P_\mu(X)\equiv\P\{X_\mu\ge X\}\le e^{- X^2/2R_0^2}.
\label{cheb}\end{equation}
Relations (\ref{ti-sa}), (\ref{ti-sa.1}) and (\ref{ti-sa.2}) prove the
self-averaging property of $\frac{1}{N}\log_{(MN)}\Theta_{N,p}(k,h,z)$.

Now let us prove that  $\ti\Delta^{(\mu)}$, defined in (\ref{t3.5}),
for any $\mu$ satisfies the bound
\begin{equation}\begin{array}{c}
|E\{\ti\Delta^{(\mu)}\}|=
|E\{\theta(Z_{N,p}^{(\mu,\mu)}-e^{-MN})
[\log_{(MN)}\lla\fH((k-x^{(\mu)})\e^{-1/2})\rra_\mu\\
-\log_{(MN)}\lla\theta(x^{(\mu)}-k)\rra_\mu]\}\le
\e^{\l}K,
\end{array}\label{t3.5a}\end{equation}
with some positive $N,\mu,\e$-independent $\l,K$. We remark here, that
similarly to (\ref{ti-sa.1a}) $Z^{(\mu-1)}_{N,p},Z^{\mu}_{N,p}\le 
Z^{\mu,\mu}_{N,p}$ and so, if $Z^{\mu,\mu}_{N,p}<e^{-MN}$, then
$\log_{(MN)}Z^{(\mu-1)}_{N,p}=\log_{(MN)}Z^{(\mu)}_{N,p}=MN$.

Using the inequalities
\begin{equation}
 \fH(-\e^{-1/4})\theta(x-\e^{1/4})\le
 \fH\left(-\frac{x}{\e^{1/2}}\right)\le \e_1+ \theta(x+\e^{1/4})
 \label{t3.6a}\end{equation}
with $\e_1\equiv \fH(\e^{-1/4})$, we get
\begin{equation}\begin{array}{c}
\log\fH(-\e^{-1/4})-E\{\theta(Z_{N,p}^{(\mu,\mu)}-e^{-MN})\log(1+r_1(k,\e))\}\\
\le E\{\tilde\Delta^{(\mu)}\}\le E\{\theta(Z_{N,p}^{(\mu,\mu)}-e^{-MN})
\log(1+r_2(k,\e))\},
\end{array}\label{t3.6}\end{equation}
where
$$\begin{array}{c}\ds{
r_1(k,\e)\equiv \frac{T_\mu(k)-T_\mu(k+\e^{1/4})}{T_\mu(k+\e^{1/4})},\quad
r_2(k,\e)\equiv\frac{T_\mu(k-\e^{1/4})-T_\mu(k)+\e_1}{T_\mu(k)}.}
\end{array}$$
But by the virtue of Lemma \ref{lem:4}, one can get easily that,
if $|X_\mu|\le \log N$,  then with probability $P_N^{(\mu)}\ge
(1-K_1N^{-3/2})$
$$
r_{1,2}(k,\e)\le \e^{1/4}Ce^{CX_\mu^2}
$$
with some $N,\mu$-independent $C$. Therefore,
choosing $\l\equiv{1\over 8}R_0^2(1+2CR_0^2)^{-1}$ and
$ L^2\equiv 2\l|\log\e|$, for small enough $\e$ we can write
similarly to (\ref{ti-sa.2})
$$\begin{array}{c}
E\left\{\theta(Z_{N,p}^{(\mu,\mu)}-e^{-MN})
\log_{(MN)}\left(1+r_{1,2}(k,\e)\right)\right\}\le (MN)P_\mu(\log N)\\
 +K_1N^{-3/2}(MN)+\int\theta(\log N-|X|)\log(1+\e^{1/4}Ce^{CX^2})dP_\mu(X)\\
=\e^{1/4}Ce^{CL^2}+C\int\theta(|X|-L)X^2dP_\mu(X)+o(1)\\
\le
\e^{1/4}Ce^{CL^2}+2CL^2P(L)\le K(C,R_0)\e^{\l},
\end{array}$$
where $P_\mu(X)$ is defined and estimated in (\ref{cheb}) and
 we have used  that, according to definition (\ref{log_M}),
 $-MN\le\log_{(MN)}\theta\la(x^{(\mu)}-k)\ra_\mu,
 \log_{(MN)}\la\theta(x^{(\mu)}-k\pm\e^{1/4})\ra_\mu\le 0$ and therefore
always $|\log_{(MN)}(1+r_{1,2}(k,\e)|\le MN$.

\smallskip

Using the bound
$$
|\frac{1}{N}\log_{(MN)}\Theta_{N,p}(k,h,z)-
\frac{1}{N}\log_{(MN)}\Theta_{N,p}(k,0,z)|\le 2hR_0,
$$
representation (\ref{t3.5}) and  self-averaging property of 
$\frac{1}{N}\log_{(MN)}\Theta_{N,p}(k,h,z)$,
we obtain that with probability $P_N\ge 1-O(N^{-1/2})$
$$\begin{array}{c}\ds{
F(\a,k,0,z,\e)+O(\e^{\l})+O(h)\le
\frac{1}{N}\log_{(MN)}\Theta_{N,p}(k,0,z)}\\
\ds{
\le F(\a,k,0,z,\e)+O(\e^{\l})+O(h)}.
\end{array}$$
Now we are going to use Corollary \ref{cor:1} to replace the integration
over the whole space by the integration over the sphere of the radius
$N^{1/2}$. But since Theorem \ref{thm:2} is valid only for $z<\e^{-1/3}$,
we need to check, that $\min_z\{F(\a,k,0,z,\e)+\frac{z}{2}\}$ takes place
for $z$, satisfying this bound.
\begin{proposition}\label{pro:3a}
For any $\alpha<\alpha_c(k)$ there exists $\e$-independent
$\ov z(k,\alpha)$ such that $z_{min}<\ov z(k,\alpha)$.
\end{proposition}

Then, using \ref{c.1}, we have got
that with the same probability for $\a\le\a_c(k)$
\begin{equation}\begin{array}{c}\ds{
\min_z\{F(\a,k,0,z,\e)+\frac{z}{2}\}
+O(\e^{\l})+O(h)\le
\frac{1}{N}\log_{(MN)}\Theta_{N,p}(k)}\\
\ds{
\le \min_z\{F(\a,k,0,z,\e)+\frac{z}{2}\}
+O(\e^{\l})+O(\de)+O(h)}.
\end{array}\label{t3.8}\end{equation}
Thus,
\begin{equation}
\lim_{N\to\infty}E\left\{\left(\frac{1}{N}\log_{(MN)}\Theta_{N,p}(k)-
E\{\frac{1}{N}\log_{(MN)}\Theta_{N,p}(k)\}\right)^2\right\}\le
O(\e^{2\l})+O(h),
\label{t3.9}\end{equation}
and since $\e,h$ are arbitrarily small numbers (\ref{t3.9}) proves the
self-averaging property of $\frac{1}{N}\log_{(MN)}\Theta_{N,p}(k)$. Besides,
averaging  $\frac{1}{N}\log_{(MN)}\Theta_{N,p}(k)$ with respect to all random variables
and taking the limits $h,\e\to 0$, we obtain
 (\ref{t2.1}) from (\ref{t3.9}).

The last statement of Theorem \ref{thm:3} follows from that proven above,
if we note that $\log_{(MN)}\Theta_{N,p}(k)$
is a monotonically decreasing function of $\alpha$ and, on the other hand,
the r.h.s. of (\ref{t3.1}) tends to $-\infty$ as $\alpha\to\alpha_c(k)$
 
  Hence, we have finished the proof of Theorem \ref{thm:3}.
 
\section{Auxiliary Results}
\no{\it Proof of Proposition \ref{pro:1}}
Let us fix $t\in(t_1^*,t_2^*)$ take some small enough $\de$ and consider
 ${\cal D}^\delta(t)$ which is the set of all $\bJ\in {\cal A}(t)\cap
{\cal D}$ whose distance from the boundary of ${\cal D}$ is more
than $d=N^{1/2}\max\{\delta,2K_0\de\}$. Now for any $\bJ_0\in
{\cal D}^\delta(t)$ consider $(\ti{\bJ},\phi(\ti{\bJ}))$ - the
local parametrisation of ${\cal D}$ with the points of the
$(N-1)$-dimensional hyper-plane ${\cal
B}=\{\ti{\bJ}:(\ti{\bJ},\ti{\bf n})=0\}$, where $\ti{\bf n}$ is
the projection of the normal ${\bf n}$ to ${\cal D}$ at the point
$\bJ_0$ on the hyper-plane ${\cal B}(t)$. We chose the orthogonal
coordinate system in ${\cal B}$  in such a way that $\ti
J_{1}=({\bJ},{\bf e})=N^{1/2}t$. Denote $\ti{\bJ_0}=P\bJ_0$ ($P$ is
the operator of the orthogonal projection on ${\cal B}$).
According to the standard theory of the Minkowski sum (see
e.g.\cite{Had}),
 the boundary of $\frac{1}{2} {\cal A}(t)\times \frac{1}{2}{\cal A}(t+\de)$
consists of the points
\begin{equation}
\bJ'=\frac{1}{2}\bJ+\frac{1}{2}\bJ^{(\de)}(\bJ),
\label{p1.1}\end{equation}
 where $\bJ$ belongs to
the boundary of  ${\cal A}(t)$  and the point
$\bJ^{(\de)}(\bJ)$ (belonging to the boundary of  ${\cal
A}(t+\delta)$) is chosen in such a way that the normal to the
boundary of  ${\cal A}(t+\de)$ at this point coincides with the
normal $\bf n$ to the boundary of  ${\cal A}(t)$ at the point
$\bJ$. Denote $\tilde{\cal D}(\frac{1}{2})$ the part of the
boundary of $\frac{1}{2} {\cal A}(t)\times \frac{1}{2}{\cal
A}(t+\de)$ for which in representation (\ref{p1.1}) $\bJ\in{\cal
D}^\delta(t)$. Now for  $\bJ_0\in{\cal D}^\delta(t)$ let us find
the point $\bJ^{(\delta)}(\bJ_0)$. Since  by construction
${\d\over\d\ti J_i}\phi(\ti{\bJ}_0)=0$ ($i=2,\dots, N-1$), we
obtain for
 $\ti{\bJ}^{(\de)}(\bJ_0)\equiv P\bJ^{(\de)}(\bJ_0)$ the system of equations
$$
{\d\over\d\ti J_i}\phi(\ti{\bJ}^{(\de)})=0,\,\, (i=2,\dots, N-1)
$$
and $\ti J^{(\delta)}_{1}=N^{1/2}(t+\de)$.
Then we get
\begin{equation}\begin{array}{c}
\ti J^{(\de)}_i=\ti J^0_i+
\de N^{1/2}(D^{-1}_{11})^{-1}(D^{-1})_{i,1}+o(\de)\,\,(i=2,\dots, N-1),
\end{array}\label{p1.2}\end{equation}
 where the matrix $\{D_{i,j}\}_{i,j=1}^{N-1}$ consists of the second
derivatives of the function $\phi (\ti{\bJ})$
($D_{i,j}\equiv\frac{\d^2}{\d \ti J_i\d \ti J_j}\phi(\ti{\bJ})$).
Thus, it was mentioned above, the point
$\bJ_1\equiv ({1\over 2}(\ti{\bJ}_0+\ti{\bJ}^{(\de)}),
{1\over 2}(\phi(\ti{\bJ}_0))+\phi(\ti{\bJ}^{(\de)}))\in
\tilde{\cal D}(\frac{1}{2})$.
Consider also the point
$\bJ_1'\equiv ({1\over 2}(\ti{\bJ}_0+\ti{\bJ}^{(\de)}),
\phi({1\over 2}(\ti{\bJ}_0+\ti{\bJ}^{(\de)})))\in
{\cal A}(t+\frac{1}{2}\delta)\cap {\cal D}$.
Then,
$$\begin{array}{c}
|\bJ_1-\bJ_1'|=
\phi\left({1\over 2}(\ti{\bJ}_0+\ti{\bJ}^{(\de)})\right)-
{1\over 2}\left(\phi(\ti{\bJ}_0)+\phi(\ti{\bJ}^{(\de)})\right)=\\
\frac{\de^2}{2}N\left((D^{-1}_{1,1})^2\sum_{i,j=2}^{N-1}
 D_{i,j}D^{-1}_{i,1}D^{-1}_{j,1}
+2D^{-1}_{1,1}\sum_{i=2}^{N-1}D_{i,1}D^{-1}_{i,1}
+D_{1,1}\right)+o(\de^2)\\
=N\de^2(D^{-1}_{1,1})^{-1}+o(\de^2).
\end{array}$$
But $(D^{-1}_{1,1})^{-1}\ge\l_{min}$, where $\l_{min}$ is the
minimal eigenvalue of the matrix $D$. Therefore, since
\begin{equation}
\l_{min}=\min_{(\ti{\bJ},\ti{\bJ})=1}(D\ti{\bJ},\ti{\bJ})\ge
\min_{(\ti{\bJ},\ti{\bJ})=1}
\frac{(D\ti{\bJ},\ti{\bJ})}{(1+\ti J_1^2({\bf n},{\bf e})^2)^{3/2}}\ge
\kappa_{min}\ge K_0N^{-1/2},
\label{p1.3}\end{equation}
we obtain that
\begin{equation}
|\bJ_1-\bJ_1'|\ge \de^2 K_0N^{1/2}.
\label{p1.4}\end{equation}
Besides, since by construction  ${\d\over\d\ti J_i} \phi(\ti{\bJ}_0)=0$ and
${\d\over\d\ti J_i}\phi(\ti{\bJ}^{(\de)})=0$, we get that the tangent
hyper-plane of the boundary $\frac{1}{2} {\cal A}(t)\times
\frac{1}{2}{\cal A}(t+\de)$  at the point $\bJ_1$ is
orthogonal to $(\bJ_1-\bJ_1')$. So, in fact, we have proved that the distance
between ${\cal D}^{\delta}(t+\frac{1}{2}\delta)$ and
$\tilde{\cal D}(\frac{1}{2})$
is more than $\de^2 K_0N^{1/2}$.
 Thus, denoting by $\tilde S(\frac{1}{2})\equiv
\hbox{mes}\tilde{\cal D}(\frac{1}{2})$, we obtain that
\begin{equation}
V(t+\frac{1}{2}\de)- \ti V(\frac{1}{2})\ge\de^2 N^{1/2} K_0
\ti S(\frac{1}{2})+o(\de^2)=\de^2 N^{1/2} K_0 S(t)+o(\de^2).
\label{p1.5}\end{equation}
Here we have used that
$\ti S(\frac{1}{2})=S(t)+o(1)$, as $\delta\to 0$,
because  the boundary ${\cal D}$ is smooth. Therefore, denoting
$\tilde V(\tau)$ the volume of $\tau {\cal A}(t)\times
(1-\tau){\cal A}(t+\de)$ and using (\ref{p1.5}), we get
$$\begin{array}{c}\ds{
 2V^{1/N}(t+\frac{1}{2}\de)-V^{1/N}(t)-V^{1/N}(t+\de)}\\
\ds{
\ge 2\left(\ti V(\frac{1}{2})+\de^2 N^{-1/2} K_0 S(t)\right)^{1/N}-
\ti V^{1/N}(0)-\ti V^{1/N}(1)+o(\de^2)}\\
\ds{
=2\ti V^{1/N}(\frac{1}{2})-\ti V^{1/N}(0)-\ti V^{1/N}(1)+
\frac{2\de^2 K_0 S(t)}{N^{1/2}\ti V^{1-1/N}(\frac{1}{2})}
+o(\de^2)}\\
\ds{
\ge \frac{2\de^2 K_0 S(t)}{N^{1/2} V^{1-1/N}(t+\frac{1}{2}\de)}+o(\de^2)
= 2\de^2  K_0C(t)V^{1/N}(t)+o(\de^2).}
\end{array}$$
Here we have used the inequality
$2\ti V^{1/N}(\frac{1}{2})-\ti V^{1/N}(0)-\ti V^{1/N}(1)\ge 0$,
which follows from the Brunn-Minkowski theorem and the relation
$V(t+\frac{1}{2}\de)=V(t)+o(1)$ (as $\de\to 0$).
Then, sending $\de\to 0$, we obtain  the statement of Proposition \ref{pro:1}.

\bigskip

\no{\it Proof of Lemma \ref{lem:1}}
Since $\log \fH(x)$ is a concave function
of $x$, $\H_{N,p}(\bJ,h,z,\e)$ is the convex function of $\bJ$,
satisfying (\ref{cond1}).
Since  $\log \fH(x)<0$ for any $x$, (\ref{cond2}) is also fulfilled.
To prove (\ref{cond3}) let us write
\begin{equation}\begin{array}{c}\ds{
|\nabla\H_{N,p}(\bJ)|^2\le\frac{3}{N\e}\sum_{i,\mu,\nu}\xmi\xni A_\mu A_\nu+
3h^2({\bf h}{\bf h})+3z^2(\bJ,\bJ)}\\
\ds{
\le\const\e^{-1} \left[\sum_{\mu}A_\mu^2+z^2(\bJ,\bJ)+h^2({\bf h}{\bf h})
\right]}\\
\ds{
\le\const\e^{-1}\left[pC^*-\sum_{\mu}\log \fH\left(k-\frac{N^{-1/2}(\bJ,\bxm)}
{\sqrt \e}\right)+h^2+z^2(\bJ,\bJ)\right]},
\end{array}\label{l1.1}\end{equation}
where we denote for simplicity
$ A_\mu\equiv \fA\left(k-\frac{N^{-1/2}(\bJ,\bxm)}{\sqrt \e}\right),$
with the function $A(x)$  defined in (\ref{A}).
The second inequality  in (\ref{l1.1}) is based on the first line of (\ref{matr_X}),
the third inequality is valid by the virtue of the bound
$\frac{1}{2}A^2(x)\le -\log H(x)+C^*$,
with some constant $C^*$, and the last inequality is valid
due to the second line of (\ref{matr_X}).

Taking into account (\ref{cond2}) one
can conclude also, that for any $U$ there exists some  $N$-independent constant
$C(U)$, such that $(\bJ,\bJ)\le NC(U)$, if $\H_{N,p}(\bJ)\le NU$.
Thus, we can derive from (\ref{l1.1}) that under conditions (\ref{matr_X})
(\ref{cond3}) is fulfilled.
 Besides, due to the inequality $\log H(x)\ge
C_1^*-{1\over 2}x^2$, it is easy to obtain that
$$
f_{N,p}(k,h,z,\e)\ge C_1^*+ {1\over N}\log \hbox{det}(\e^{-2}X+zI),
$$
so (\ref{cond4}) is also fulfilled.

Hence, we have proved that under conditions (\ref{matr_X}) the norm of the
matrix ${\cal D}\equiv\{\la\dot J_i\dot J_j\ra\}_{i,j=1}^N$ is bounded by some $N$-independent
 $C(z,\e)$. Then with the same probability
$$
N^{-1}\sum_{i,j=1}^N\la\dot J_i\dot J_j\ra^2=
N^{-1}\hbox{Tr}{\cal D}^2\le C(z,\e),
$$
which implies (\ref{main}).

\smallskip

To prove (\ref{bound_J}) let us observe that
\begin{equation}
\la\theta(|J_N|-N^{1/2}\e_N) \ra=\la\theta(|c|-\e_N)\ra_{(U,c)},
\label{l1.1a}\end{equation}
where $\la\dots\ra_{(U,c)}$ is defined in (\ref{t1.5})-
(\ref{t1.8a}) with ${\bf e}=(0,\dots,0,1)$. For the
function $s_N(U,c)$, defined by (\ref{t1.7}), we get
\begin{equation}\begin{array}{c}\ds{
\la \frac{\partial}{\partial c} s_N(U,0)\ra_{(U,0)}=
N^{-1/2}\frac{\int\frac{\partial}{\partial J_N}
{\cal H}_{N.p}(\bJ)\exp\{-{\cal
H}_{N.p}(\bJ)\}|_{J_N=0}d J_1\dots dJ_{N-1}}
{\exp\{-{\cal
H}_{N.p}(\bJ)\}|_{J_N=0}d J_1\dots dJ_{N-1}}}\\
\ds{
=\frac{hh_N}{N^{1/2}}+\frac{1}{N\e}\sum_{\mu=1}^p\xmN\la
A_\mu\ra\bigg|_{J_N=0}.}
\end{array}\label{l1.2}\end{equation}
But since $\la A_\mu\ra|_{J_N=0}$ does not depend on $\xmN$,
by using the standard Chebyshev inequality, we obtain that
\begin{equation}
\P\left\{|\la \frac{\partial}{\partial c} s_N(U,0)\ra_{(U,0)}|
>\e_N\right\}\le  e^{-C_1 N\e_N^2}=e^{-C_1\log^2N}.
\label{l1.3}\end{equation}
On the other hand, since $s_N(U,c)$ is a concave function of $U,c$
satisfying (\ref{t1.10a}), denoting $\phi_N(U,c)\equiv 
s_N(U,c)-U-(s_N(U^*,c^*)-U^*)$ for any $(U,c)\sim (U^*,c^*)$ one can write
\begin{equation}
C_0[(U-U^*)^2+(c-c^*)^2]\le -\frac{\partial}{\partial c}\phi_N(U,c)(c-c^*)
 -\frac{\partial}{\partial U}\phi_N(U,c)(U-U^*).
\label{l1.4}\end{equation} 
Multiplying this inequality by $e^{N \phi_N(U,c)}$ and integrating with respect
to $U$, we obtain for $c=0$
$$\begin{array}{c}\ds{
C_0(c^*)^2\le c^*\la \frac{\partial}{\partial c} s_N(U,0)\ra_{(U,0)} +O(N^{-1}).}
\end{array}$$
Therefore, taking into account (\ref{l1.3}), we get that, if
\ref{matr_X} is fulfilled, then
\begin{equation}
\P\left\{|c^*|
>\frac{\e_N}{2}\right\}\le e^{-C_1\log^2N}.
\label{l1.5}\end{equation}
But, using the Laplace method, we get easily
$$
\la\theta(|c-c^*|-\frac{\e_N}{2})\ra_{(U,c)}\le e^{-CN\e_N^2}\le
e^{-C\log^2N}.
$$
Combining this inequality with (\ref{l1.1a}) and using the symmetry with respect
to $J_1,\dots,J_N$, we obtain (\ref{bound_J}).

\medskip

\no{\it Proof of Proposition \ref{pro:2}}
Applying Lemma \ref{lem:2} to the sequences $f_{N_m,p_m}$ and
$f_{N_m,p_m-1}$ as a functions
of $z$,  we obtain immediately relations (\ref{p2.1}) for $R_{N_m,p_m}$
for all $z$, where the limiting free energy $f(z,h)$
 has continuous first derivative with respect to $z$. Besides, since
 for all $\lambda\in (-1,1)$ and arbitrarily small $\de>0$
 $$\begin{array}{c}
\lambda E\left\{\de^{-1}\left(f_{N_m,p_m}(z-\de)-f_{N_m,p_m}(z-2\de)\right)\right\}\le
 E\left\{\log\lla\exp\left\{\lambda N_m^{-1}(\bJ,\bJ)\right\}\rra\right\}\\
\le \lambda E\left\{ \left(\de^{-1}(f_{N_m,p_m}(z+2\de)-f_{N_m,p_m}(z+\de)\right)\right\},
 \end{array}$$
we obtain that $E\left\{\log\lla\exp\{\lambda(N_m^{-1}(\bJ,\bJ)\}\rra-
\ov R_{N_m,p_m}) \right\}\to 0$
for all such $z$ and all $\lambda\in (-1,1)$. Using Remark \ref{rem:3}, we can derive 
then that 
$$
f_m(\lambda)\equiv E\left\{ \lla\exp\left\{\lambda(N_m^{-1}(\bJ,\bJ)-
\ov R_{N_m,p_m})\right\}\rra
\right\}\to 1.
$$
Then, since it follows from Remark \ref{rem:3} that $f_k^{(3)}(\lambda)$ is 
bounded uniformly in $m$ and $\lambda$, we derive that 
$f_m''(\lambda)\to 0$ and, taking here $\lambda=0$, obtain (\ref{s-a.R}).

To derive relations (\ref{p2.1}) for $q_{N_m,p_m}$
 we consider $f_{N_m,p_m}$ and
$f_{N_m,p_m-1}$ as a functions of $h$, derive from Lemma \ref{lem:2} that
$$
E\left\{\left(N_m^{-1}({\bf h},\la \bJ\ra_{N_m,p_m})-E\left\{
N_m^{-1}({\bf h},\la \bJ\ra_{N_m,p_m})\right\}\right)^2\right\}\to 0
$$
and therefore
$$\begin{array}{c}\ds{
E\left\{\left(N_m^{-1}({\bf h},\la \bJ\ra_{N_m,p_m})-E\left\{
N_m^{-1}({\bf h},\la \bJ\ra_{N_m,p_m})\right\}\right)
N_m^{-1}(\la \bJ\ra_{N_m,p_m},\la \bJ\ra_{N_m,p_m})\right\}\to 0}.
\end{array}$$
Integrating it with respect to $h_i$, we get
$$\begin{array}{c}
E\left\{ (q_{N_m,p_m}-\ov q_{N_m,p_m}-(R_{N_m,p_m}-\ov R_{N_m,p_m}))
q_{N_m,p_m}\right\}\\
\ds{
={2\over N_m^2}\sum_{i,j=1}^{N_m}E\left\{\la J_i\ra\la\dot J_i\dot J_j\ra
\la J_j\ra\right\}.}
\end{array}$$
Using  relations (\ref{main}) and (\ref{s-a.4}) we derive now
(\ref{p2.1}) for $q_{N_m,p_m}$.

\bigskip

\no{\it Proof of Lemma \ref{lem:3}}
Let us note that, by the virtue of Lemma \ref{lem:1}, computing
$\phi_N(\e_1,k_1)$, $\phi_{0,N}(\e_1,k_1)$  with  probability
more than $(1-e^{-C_2\log^4N})$ we
can restrict all the integrals with respect to $\bJ$ by the domain
$$
\Omega_N=\left\{|J_i|\le\e_NN^{1/2},\,\,(i=1,\dots,N),(\bJ,\bJ)
\le N R_0^2\right\}
$$
In this case the error for $\phi_N(\e_1,k_1)$ and $\phi_{0,N}(\e_1,k_1)$
will be  of the order $O(Ne^{-C_1\log^2N})$. So below in the proof of
Lemma \ref{lem:3} we  denote by $\la...\ra_{p-1}$ the Gibbs measure,
corresponding to the Hamiltonian $H_{N,p-1}$ in the domain $\Omega_N$.
In this case the inequalities (\ref{main}) are also valid, because
their l.h.s., comparing with those, computing in the whole ${\bf R}^N$,
have the errors of the order $O(N^2e^{-C_1\log^2N})$.

\smallskip

 We start from the proof of the first line
of (\ref{l3.3}). To this end consider the functions
\begin{equation}\begin{array}{l}
F_N(t)\equiv\lla\theta(N^{-1/2}(\bxm,\bJ)-t)\rra_{p-1},\\
F_{0,N}(t)=\fH\left(  U_{N,p}^{-1/2}(0)\left(
 N^{-1/2}(\bxm,\la\bJ\ra_{p-1})-t\right)\right),\\
\psi_N(u)\equiv\lla\exp
\left\{ iu(\bxm,\dot{\bJ})N^{-1/2} \right\}
\rra_{p-1},\\
\psi_{0,N}(u)\equiv\exp\left\{-\frac{u^2}{2}(R_{N,p-1}-q_{N,p-1})\right\}.
\end{array}\label{ti-phi}\end{equation}
Take $L\equiv{\pi\over 4\e_N}$. According to the Lyapunov theorem 
(see \cite{Lo}),
\begin{equation}
\max_{t}|F_N(t)-F_{0,N}(t)|\le{2\over\pi}\int_{-L}^Lu^{-1}du|\psi_N(u)-
\psi_{0,N}(u)|+\frac{\const}{L}.
\label{Lyap}\end{equation}
Since evidently
$$\begin{array}{c}
\phi_N(\e_1,k_1)=\e_1^{1/2}\int \fH(\e_1^{-1/2}(k_1-t))dF_N(t),\\
\phi_{0,N}(\e_1,k_1)=\e_1^{1/2}\int \fH(\e_1^{-1/2}(k_1-t))dF_{0,N}(t),
\end{array}$$
we obtain
\begin{equation}
|\phi_N(\e_1,k_1)-\phi_{0,N}(\e_1,k_1)|\le\max_t|F_N(t)-F_{0,N}(t)|\const.
\label{l3.41}\end{equation}
Thus, using (\ref{Lyap}), we have got
\begin{equation}\begin{array}{c}\ds{
E\left\{|\phi_N(\e_1,k_1)-\phi_{0,N}(\e_1,k_1)|^2\right\}\le
\const({1\over L}+I_1+I_2),}\\
\ds{
I_1\equiv E\left\{\int_1^1 u^{-2}|\psi_N(u)-\psi_{0,N}(u)|^2du\right\},}\\
\ds{
I_2\equiv \int_{1<|u^{(1)}|,|u^{(2)}|<L}du^{(1)}du^{(2)}E\left\{
(\psi_N(u^{(1)})-\psi_{0,N}(u^{(1)}))\right.}\\
\left.\cdot(\ov\psi_N(u^{(2)})-
\ov\psi_{0,N}(u^{(2)}))\right\}.
\end{array}\label{l3.42}\end{equation}

 Consider
\begin{equation}\begin{array}{c}\ds{
I_2^{(1)}\equiv E_p\left\{\int_{1<|u^{(1)}|,|u^{(2)}|<L}du^{(1)}du^{(2)}\psi_N(u^{(1)})
\ov\psi_N(u^{(2)})\right\}}\\
=\int_{1<|u^{(1)}|,|u^{(2)}|<L}du^{(1)}du^{(2)}
\lla\prod_{j=1}^N\cos N^{-1/2}
\left(u^{(1)}\dot J_j^{(1)}-u^{(2)}\dot J_j^{(2)}\right)\rra_{p-1}.
\end{array}\label{l3.5}\end{equation}
 We would like to prove  that one can substitute the product of
$\cos(a_i)$ in (\ref{l3.5}) by the product of $\exp\{-a_i^2/2\}$.
So we should estimate
\begin{equation}\begin{array}{c}
\Delta\equiv E\left\{\int_{1<|u^{(1)}|,|u^{(2)}|<L} du^{(1)} du^{(2)}
\lla\left[\prod_{j=1}^N\cos N^{-1/2}
\left(u^{(1)}\dot J_j^{(1)}-u^{(2)}\dot J_j^{(2)}\right)\right.
\right.\right.\\
\left.\left.\left.-\exp\left\{-{1\over 2N}\sum \left(u^{(1)}\dot J_j^{(1)}-
u^{(2)}\dot J_j^{(2)}\right)^2\right\}\right]\rra_{p-1}\right\}.
\end{array}\label{l3.6}\end{equation}
Let us denote
$$\begin{array}{c}\ds{
g(\tau)\equiv\sum_i\left(\log\cos N^{-1/2}\tau\left(u^{(1)}\dot J_j^{(1)}
-u^{(2)}\dot J_j^{(2)}\right)\right.}\\
\ds{
\left.+{\tau^2\over 2N}\sum\left(u^{(1)}J_j^{(1)}-u^{(2)}J_j^{(2)}\right)^2\right).}
\end{array}$$
Then
\begin{equation}\begin{array}{c}\ds{
|\Delta|=\left|\int_{1<|u^{(1)}|,|u^{(2)}|<L} du^{(1)} du^{(2)}
\la e^{g(1)}-e^{g(0)}\ra\right| }\\
\ds{
\le\int_{|u^{(1)}|,|u^{(2)}|<L} du^{(1)} du^{(2)}\lla
|g(1)-g(o)|(e^{g(1)}+e^{g(0)})\right.}\\
\left.\exp\left\{-{1\over 2N}
\sum \left(u^{(1)}\dot J_j^{(1)}-u^{(2)}\dot J_j^{(2)}\right)^2\right\}
\rra_{p-1}.
\end{array}\label{l3.7}\end{equation}
But since $g(0),g'(0),g''(0),g'''(0)=0$,
$$\begin{array}{c}\ds{
|g(1)-g(0)|\le {1\over 6}|g^{(4)}(\zeta)|\le{\const \over N^{2}}
\sum \left(u^{(1)}\dot J_j^{(1)}+u^{(2)}\dot J_j^{(2)}\right)^4}\\
\ds{
\le \const \e_N^2\left[\left(N^{-1}(\dot{\bJ}^{(1)},
\dot{\bJ}^{(1)})+N^{-1}(\dot{\bJ}^{(2)},\dot{\bJ}^{(2)})\right)
\left(|u^{(1)}|^4+|u^{(2)}|^4\right)\right].}
\end{array}$$
Besides, using the inequality (valid for any $|x|\le\frac{\pi}{2}$)
$$
\log\cos x+\frac{x^2}{2}\le \frac{x^2}{6},
$$
we obtain that
$$
|e^{g(0)}+e^{g(1)}|\le 2\exp\left\{\frac{1}{6N}
\sum \left(u^{(1)}\dot J_j^{(1)}+u^{(2)}\dot J_j^{(2)}\right)^2\right\}.
$$
Thus, we get from (\ref{l3.7})  $|\Delta|\le\const\e_N^2$.
Hence, we have proved  that
\begin{equation}\begin{array}{c}\ds{
I_2^{(1)}
=\int du^{(1)} du^{(2)}\lla\exp\left\{-{1\over 2}\sum_{l,m=1}^2
A^{(1)}_{l,m}u^{(l)}u^{(m)}\right\}
\rra^{(1,2)}_{p-1}+O(\e_N^2),}
\end{array}\label{l3.8}\end{equation}
where
$$
 A^{(1)}_{l,l}={1\over N}(\dot{\bJ^{(l)}},\dot{\bJ^{(l)}}),\,\,\ (l=1,2)\,\,\,
 A^{(1)}_{1,2}={1\over N}(\dot{\bJ^{(1)}},\dot{\bJ^{(2)}}).
$$
Now, taking into account that  Proposition \ref{pro:2} implies
$$
 \sum_{m,l=1,2}E\left\{\lla( A^{(1)}_{l,m}- A_{l,m})^2
\rra^{(1,2)}_{p-1}\right\}\to
0,\,\,( N\to\infty),
$$
where $A_{l,m}=\delta_{l,m} (R_{N,p-1}-q_{N,p-1}),$
we obtain immediately that
$$\begin{array}{c}\ds{
\int_{1<|u^{(1)}|,|u^{(2)}|<L}du^{(1)}du^{(2)}
E\left\{\psi_N(u^{(1)})\ov\psi_N(u^{(2)})\right\}}\\
\ds{
=\int_{1<|u^{(1)}|,|u^{(2)}|<L}du^{(1)}du^{(2)}
E\left\{\psi_{0,N}(u^{(1)})\ov\psi_{0,N}(u^{(2)})\right\}+o(1).}
\end{array}$$
By the same way one can prove also
$$\begin{array}{c}\ds{
\Re\int_{1<|u^{(1)}|,|u^{(2)}|<L}du^{(1)}du^{(2)}
E\left\{\psi_N(u^{(1)})\ov\psi_{0,N}(u^{(2)})\right\}}\\
\ds{
=\int_{1<|u^{(1)}|,|u^{(2)}|<L}du^{(1)}du^{(2)}
E\left\{\psi_{0,N}(u^{(1)})\ov\psi_{0,N}(u^{(2)})\right\}+o(1)},
\end{array}$$
which gives us that $I_2=o(1)$. Similarly one can prove that
$I_1=o(1)$. Then,  using
(\ref{l3.42}), we obtain the first line of (\ref{l3.3}).

To  prove the second line of (\ref{l3.3}) we denote by
$A\equiv(\phi_{N}(\e_1,k_1))$,  $B\equiv(\phi_{0,N}(\e_1,k_1))$,
$\ti\e_N\equiv E\{(A-B)^2\}$, $\ti L\equiv|\log\ti\e_N|\ti\e_N^{-1/2}$
and write
\begin{equation}\begin{array}{c}
E\left\{|\log A-\log B|^2\right\}\le E\left\{\theta(\ti L-A^{-1})
\theta(\ti L-B^{-1})
(|\log A-\log B|^2\right\}\\
+2E\left\{(\theta(\ti L-A^{-1})+\theta(\ti L-B^{-1})))
(\log^2 A+\log^2 B)\right\}\le \\
4\ti L^{-2}E\left\{
( A- B)^2\right\}+4|\log\ti L|^{-2} E\left\{(\log^4 A+\log^4 B)\right\}\\
\le 4\ti\e_N\ti L^{-2}+|\log\ti L|^{-2}\const\le\const |\log\ti L|^{-3/2}.
\end{array}\label{l3.10}\end{equation}
Here we have used the inequality
$$
|\log A-\log B|\le |A-B|(A^{-1}+B^{-1}),
$$
the first line of (\ref{l3.3}) and the fact that $E\{\log^4A\}$,
$E\{\log^4B\}$ are bounded (it can be obtained similarly to 
(\ref{s-a.5})-(\ref{s-a.6})).
Since we have proved above that $\ti\e_N\to 0$, as $N\to\infty$,
inequality (\ref{l3.10}) implies the second line of (\ref{l3.3}).
The third and the fourth line of (\ref{l3.3}) can be derived in the usual way
(see e.g. \cite{PST2}) from
the second line by using the fact that functions $\log\phi_{N}(\e_1,k_1)$
and $\log\phi_{0,N}(\e_1,k_1)$ are convex with respect to $\e_1^{-1}$
and $k_1$.

The convergence in  distribution
$N^{-1/2}(\bxp,\la \bJ\ra_{p-1})\to
\sqrt{\ov q_{N,p}}u$ follows from the central limit theorem (see, e.g.
the book \cite{Lo}), because $\la \bJ\ra_{p-1}$ does not depend on $\bxp$
and the Lindenberg condition is fulfilled:
$$
{1\over N^2}\sum_i\la J_i\ra_{p-1}^4\le\const \e_N^2.
$$
\medskip

Thus, to finish the proof of Lemma \ref{lem:3} we are left to prove
(\ref{l3.4}). It can be easily done, e.g. for $\mu=p$ and $\nu=p-1$,
if we in the same manner as above consider the functions

\begin{equation}\begin{array}{c}\ds{
\phi_N^{(2)}(\e_1,\e_2,k_1,k_2)\equiv\int_{x_1,x_2>0}dx_1dx_2\lla
\exp\left\{-{1\over 2\e_1}(N^{-1/2}
(\bxp,\bJ)-x_1-k_1)^2\right.\right.}\\
\ds{
\left.\left.-{1\over 2\e_2}(N^{-1/2}(\bxp,\bJ)-x_2-k_2)^2\right\}\rra_{p-1}}
\end{array}\label{l3.12}\end{equation}
\begin{equation}\begin{array}{c}\ds{
\phi_{0,N}^{(2)}(\e_1,\e_2,k_1,k_2)}\\
\ds{
\equiv(\e_1\e_2)^{1/2}\fH\biggl({N^{-1/2}(\bxp,\la \bJ\ra_{p-2})
-k_1\over U_{N,p-2}^{1/2}(\e_1)}\biggr)
\fH\biggl({N^{-1/2}(\mbox{\boldmath$\xi^{(p-1)}$},\la \bJ\ra_{p-2})
-k_2\over U_{N,p-2}^{1/2}(\e_2)}\biggr)}
\end{array}\label{l3.13}\end{equation}
and prove for them analog of relations (\ref{l3.3}). Then relations (\ref{l3.4})
will follow immediately. The self-averaging property for $\ti U_N$ and
$\ti q_N$ follows from the fact that $\phi_{0,N}^{(2)}(\e_1,\e_2,k_1,k_2)$
is a product of two independent functions.

\bigskip

\no{\it Proof of  Proposition \ref{pro:3}}.
It is easy to see, that equations (\ref{q,R})
have the form
\begin{equation}
{\d F\over \d q}=O(\ti\e_N),\quad {\d F\over \d R}=O(\ti\e_N'),
\label{p3.1}\end{equation}
where $F(q,R)$ is defined by the expression in the square brackets in the
r.h.s. of (\ref{t2.1}).

Let us  make the change of variables
$s=q(R+\e-q)^{-1}$. Then equations (\ref{p3.1}) take the form
\begin{equation}
{\d \ti F\over \d s}=O(\ov\e_N),\quad {\d \ti F\over \d R}=O(\ov\e_N),
\label{p3.3}\end{equation}
where $\ov\e_N=|\ti\e_N|+|\ti\e_N'|$ and
\begin{equation}\begin{array}{c}\ds{
\ti F(s,R)\equiv\a E\left\{\log \fH\left(u\sqrt s+{k\sqrt{1+s}\over
\sqrt{\e+R}}\right)
\right\}}\\
\ds{
+{1\over 2}{s(R+\e)\over R-\e s}+{1\over 2}\log(R-\e s)
-{1\over 2}\log(1+s)-{z\over 2}R+
{h^2\over 2}{R-\e s\over 1+s}.}
\end{array}\label{p3.4}\end{equation}
Then  (\ref{p3.3}) can be written in the form
\begin{equation}\begin{array}{c}\ds{
f_1(s,R)\equiv-{\a\over s} E\left\{\fA^2\right\}
+\frac{(R+\e)^2}{(R-\e s)^2}-\frac{h^2}{s(s+1)}(R+\e)=O(\ov\e_N),}\\
\ds{
f_2(s,R)\equiv\frac{\alpha k\sqrt{1+s}}{(R+\e)^{3/2}}E\left\{\fA\right\}-
\frac{\e s(s+1)}{(R-\e s)^2}+\frac{1}{R-\e s}+\frac{h^2}{s+1}-z=O(\ov\e_N),}
\end{array}\label{p3.5}\end{equation}
where the function $A(x)$ is defined by (\ref{A}) and
to simplify formulae we  here and below omit the arguments of functions
 $A$ and $A'$. But
\begin{equation}\begin{array}{c}\ds{
\frac{\d}{\d s}f_1(s,R)=-\frac{\a}{s^2}E\left\{\left(u\sqrt s
+{k\sqrt{1+s}\over\sqrt{\e+R}}\right)\fA'\fA\right\}
+\frac{\a}{s^2}E\left\{\fA^2\right\}}\\
\ds{
+\frac{\a k}{s^2(1+s)^{1/2}(\e+R)^{1/2}}E\left\{\fA'\fA\right\}+
\frac{2(R+\e)^2\e}{(R-\e s)^3}+\frac{h^2(2s+1)}{s^2(s+1)^2}(R+\e)>
\frac{h^2}{s^2}R.}
\end{array}\label{p3.6}\end{equation}
Here 
we have used the inequality (we prove it below):
\begin{equation}
x\fA'(x)\fA(x)\le A^2(x),
\label{p3.6a}\end{equation}
which gives us that the sum of the first two terms in  (\ref{p3.6}) is positive.
Therefore we conclude, that equation ${\d \ti F\over \d s}(s,R)=0$
for any $R$ has a unique solution $s=s(R)$ and, if we consider
the first of equations (\ref{p3.3}), then its solution  $s_1(R)$ for any
$R$ behave like
\begin{equation}
s_1(R)=s(R)+O(\ov\e_N).
\label{p3.9}\end{equation}
For $k=0$ the second equation in (\ref{p3.5})
 is quadratic with respect to $(R-\e s)$,
and so we can easily obtain that the system (\ref{p3.5}) for $z<\e^{-1/3}$
has the unique solution. Consider now the case, when $k\not=0$. Then
the  function $f_2(s,R)$ for $s>>1$ behaves like
$$
f_2(s,R)\sim\frac{s}{R+\e}D+\frac{1}{R-\e s}-z,
$$
where
$$\begin{array}{c}
D\equiv \alpha \ti k I_1(\ti k)-\alpha  I_2(\ti k)+
\sqrt{\alpha I_2(\ti k)},\,\, \ti k=k(R+\e)^{-1/2}\\
I_{1,2}(\ti k)\equiv\frac{1}{\sqrt{2\pi}}\int_{-\ti k}^\infty(u+\ti k)^{1,2}
e^{-u^2/2}du.
\end{array}$$
Since $D$ can be represented in the form
$$
D=\frac{\alpha \fH(-\ti k)}{\alpha \fH(-\ti k)+\sqrt{\alpha I_2(-\ti k)}}
\left[(\ti k\fA(-\ti k)+\ti k^2+1-2\fH(-\ti k))+(2-\alpha)\fH(-\ti k)\right]>0
$$
(we have checked that  $\ti k\fA(-\ti k)+\ti k^2+1-2\fH(-\ti k)\ge 0$ 
numerically),
we get from (\ref{p3.5}) that the inequality $z\le \e^{-1/3}$ implies that
$s\le \const R\e^{-1/3}$.
On the other hand,
\begin{equation}\begin{array}{c}\ds{
2{\d^2 \ti F\over \d R^2}(s,R)=
-{3\alpha k\sqrt{s+1}\over 2(R+\e)^{5/2}}E\left\{\fA\right\}
-{\a k^2 (s+1)\over 2(R+\e)^3}E\left\{\fA'\right\}
-{R-3\e s-2s^2\e\over (R-\e s)^3}.}
\end{array}\label{p3.7}\end{equation}
We would like to remark here, that  the case
when $s$ is bounded and $R\sim\e$ is impossible for $k\not=0$ due to the first
equation in (\ref{p3.5}).
Thus, for $\e$ small enough we get
\begin{equation}
{\d^2 \ti F\over \d R^2}(s,R)<-{1\over 2(R-\e s)^2}.
\label{p3.8}\end{equation}

 Now, if we consider the
function $\varphi(R)\equiv\ti F(s(R), R)$ it is obviously concave
and therefore the equation $ \varphi'(R)=0$ has the unique solution
$R^*$ which is a maximum of $\varphi(R)$. Besides, since in view of
(\ref{p3.8}) $\varphi''(R)<0$,  $R(\ti \e)$ - the solution of equation
$\varphi'(R)=\ti\e$ has the form $R(\ti \e)=R^*+O(\ti\e)$.
But in view of (\ref{p3.9}) the second equation of (\ref{p3.3})
can be rewritten in the form
$$
\varphi'(R)=O(\ti\e_N)+O(\ov\e_N).
$$
Therefore its solution tends to $R^*$ as $\ov\e_N\to 0$.

Now to finish the proof of Proposition \ref{pro:3} we are left to prove
(\ref{p3.6a}). For  $x\le 0$ it is evidently fulfilled. For $x>0$ let us write
$$\begin{array}{c}
x\fA'(x)\fA(x)=x(\fA(x)-x)\fA^2(x)\le x\frac{\sqrt{x^2+4}-x}{2}\fA^2(x)\\
\ds{
=\frac{2x}{\sqrt{x^2+4}+x}\fA^2(x)\le \fA^2(x),}
\end{array}$$
where we have used the well known  inequality (see, e.g.,\cite{Abr})
$$
\fA(x)\le \frac{\sqrt{x^2+4}+x}{2}.
$$
Proposition \ref{pro:3} is proven.

\bigskip
\no{\it Proof of Proposition \ref{pro:3a}.}

\no One can see easily that, if we want to study
$\min_z\{F(\a,k,0,z,\e)+\frac{z}{2}\}$, then we should consider the system
(\ref{p3.5}) with zeros in the r.h.s. and with the additional equation
$$
\frac{\partial}{\partial z}F(\a,k,0,z,\e)=1\Leftrightarrow R=1
$$
Thus, we need to substitute $R=1$, in the first equation. Since the
l.h.s. of this equation for $\e=0$ is an increasing function which
tends to $1-\alpha\alpha_c^{-1}>0$, as $s\to \infty$, there exist
the unique $s^*$, which is the solution of this equation. Then, choosing
$\e$ small enough, it is easy to obtain, that  $s(\e)$ is in some
$\e$-neighbourhood of $s^*$ and therefore $s(\e)\le \ov s(k,\alpha)$. Then, 
substituting this $s(\e)$ in the second equation, we get the $\e$-independent
 bound for $z$.

\bigskip

\no{\it Proof of Lemma \ref{lem:4}.}
Repeating conclusions (\ref{t1.5})-(\ref{t1.8}) of the proof of
 Theorem \ref{thm:1}, one can  see that
\begin{equation}\begin{array}{c}\ds{
\la \theta(x^{(\mu)}- k)\ra_\mu=
\la \theta(c-kN^{-1/2})\ra_{(U,c)},}
\end{array}\label{l4.3}\end{equation}
where $\la\dots\ra_{(U,c)}$
are defined by (\ref{t1.8a}) (see also (\ref{t1.5}), (\ref{t1.7})
for $\Gamma_N=\Omega_{N,p}^{(\mu-1)}$, $\Phi_N={\cal H}_{N,p}^{(\mu)}$
and $c=N^{-1}\sum\xmi J_i$.
We denote $ \phi_N^{(\mu)}(c,U)\equiv (s_N^{(\mu)}(c,U)-U-
(s_N^{(\mu)}(c^*,U^*)-U^*))$, where $s_N^{(\mu)}(c,U)$ is defined by
(\ref{t1.7}) and $(c^*,U^*)$ is the point of maximum of the function
$s_N^{(\mu)}(c,U)-U$.

Applying Theorem \ref{thm:1}, we found that $s_N^{(\mu)}(c,U)$ is a 
concave function of $(c,U)$ and it satisfies (\ref{t1.12}).

 Denote
\begin{equation}
\Lambda_M\equiv\{(U,c): N\phi_N^{(\mu)}(c,U)\ge M\},\,\
\Pi_{c^*,\ti c'} \equiv\{(U,c):c^*\le c\le\ti c'\},
\label{p4.2a}\end{equation}
and  let for any measurable ${\cal B}\subset {\bf R}^2$
 $m({\cal B})\equiv\la\chi_{\cal B}(c,U)\ra_{(U,c)}$.
 
 To prove Lemma \ref{lem:4} we use the following statement:
 
\begin{proposition}\label{pro:4} If the function
$\phi_N^{(\mu)}(c,U)$ is concave and satisfies inequality
(\ref{t1.12}),  $\ti c, \ti c'>c^*$,  and
the constant $A\le -\frac{N^{1/2}}{2(\ti c-c^*)}\max_U \phi_N^{(\mu)}(\ti c,U)$, then
\begin{equation}\begin{array}{c}\ds{
 \frac{\la \theta(c-\ti c)e^{AN^{1/2}c}\ra_{(U,c)}}
{\la\theta(c-\ti c)\ra_{(U,c)}}\le 2e^{\sqrt N A\ti c},}
\end{array}\label{p4.1}\end{equation}
and for any $M<-4$
\begin{equation}
m(\ov\Lambda_M)\le\frac{1}{4},\quad 
\frac{m(\ov\Lambda_M\cap\Pi_{c^*,\ti c'})}{m(\Lambda_M\cap\Pi_{c^*,\ti c'})}
\le \frac{1}{4}.
\label{p4.2}\end{equation}
\end{proposition}
The proof of this Proposition is given after the proof of Lemma \ref{lem:4}.

Let us choose any  $\ti c>c^*$ and
$A=-\frac{N^{1/2}}{2(\ti c-c^*)}\max_U \phi_N^{(\mu)}(\ti c,U)$. 
Using (\ref{p4.1}), we get 
\begin{equation}\begin{array}{c}
\ds{ \lla e^{AN^{1/2}(c-\ti c)}\rra_{(U,c)}=
\la\theta(c_2-c)\ra_{(U,c)}
+\frac{\la \theta(c-\ti c)e^{AN^{1/2}c}\ra_{(U,c)}}
{\la\theta(c-\ti c)\ra_{(U,c)}}\la\theta(c-c_2)\ra_{(U,c)}}\\
\ds{\le
\la\theta(c_2-c)\ra_{(U,c)}
+2\la\theta(c-c_2)\ra_{(U,c)}\le 2}
\end{array}\label{l4.4a}\end{equation}
On the other hand, we shall prove below 
\begin{proposition}\label{pro:5}
For any $|A|\le O(\log N)$
\begin{equation}\begin{array}{c}
g(A)\equiv\log\lla \exp\{AN^{1/2}(c-\la c\ra)\}
\rra_{(U,c)}=
\log\lla\exp\{AN^{-1/2}(\bxm,\dot{\bJ})\}\rra_\mu\\
= {A^2\over 2N}\lla(\dot{\bJ},\dot{\bJ})\rra_\mu+R_N,\quad
E\left\{R_N^4\right\}=O(A^{16}N^{-2}) .
\end{array}\label{l4.5}\end{equation}
\end{proposition}
It follows from this proposition that the probability to 
have for all  

\no $A_i=\pm 1,\dots,\pm[\log N]$ the inequalities
\begin{equation}
e^{A_i^2R_0^2}\ge\lla \exp\left\{A_i N^{1/2}(c-\la c\ra)\right\}\rra_{(U,c)}\ge
e^{A_i^2D^2/4} \label{l4.6}\end{equation} 
is more than $P_N'\ge 1-O(N^{-3/2})$. Therefore, using that
$\log\la\exp\{A N^{1/2}(c-\la c\ra)\}\ra_{(U,c)}$ is a convex 
function of $A$, and this function is zero for $A=0$, one can conclude 
that with the same probability
for any $A:\,\, 1\le|A|\le\log N$
\begin{equation}
e^{2A^2R_0^2}\ge\lla \exp\left\{A N^{1/2}(c-\la c\ra)\right\}\rra_{(U,c)}\ge
e^{A^2D^2/8}. 
\label{l4.6a}\end{equation} 
The first of  these inequalities implies, in particular, that for any
 $0<L<\log N$
\begin{equation}\begin{array}{c}
\la\theta(\la c\ra-LN^{-1/2}-c)\ra_{(U,c)}\\
\le \max_{A >0}\lla \exp\left\{A N^{1/2}(\la c\ra-
LN^{-1/2}-c)\right\}\rra_{(U,c)}
\le e^{-L^2/8R_0^2}.
\end{array}\label{l4.6b}\end{equation}
The same bound is valid for $\la\theta(c-\la c\ra-LN^{-1/2})\ra_{(U,c)}$.
Thus, assuming that $\la c\ra>c^*$ and denoting 
$L_0=\frac{1}{2}N^{1/2}(\la c\ra-c^*)$, $c_1\equiv\la c\ra-2L_0N^{-1/2}=c^*$, 
$c_2\equiv\la c\ra-L_0N^{-1/2}$, $c_3\equiv\la c\ra+L_0N^{-1/2}$
we can write
\begin{equation}\begin{array}{c}\ds{
1=\la\theta(c_1-c)\ra_{(U,c)}+
\la\chi_{c_1,c_3}(c)\ra_{(U,c)}
+ \la\theta(c-c_3)\ra_{(U,c)}
\le 4e^{-L_0^2/8R_0^2}.}\\
\Rightarrow  N|\la c\ra-c^*|^2=4L_0^2\le 16R_0^2.
\end{array}\label{l4.6c}\end{equation}
Here we have used (\ref{l4.6b}) and the fact that since
$\phi^{(\mu)}_N(U,c)$ is a concave function and $(U^*,c^*)$ is the point of
its maximum, we have for any $d>0$ and $\ti c>c^*$
\begin{equation}\begin{array}{c}
\la\chi_{\ti c, \ti c+d}(c)\ra_{(U,c)}\le \la\chi_{c^*,c^*+d}(c)\ra_{(U,c)}
\Rightarrow\\
\la\chi_{c_2,\la c\ra}(c)\ra_{(U,c)},
\la\chi_{\la c\ra, c_3}(c)\ra_{(U,c)}
\le\la\chi_{c^*,c_2}(c)\ra_{(U,c)}\le
\la\theta(c^*-c)\ra_{(U,c)}\le e^{-L_0^2/8R_0^2}.
\end{array}\label{l4.7}\end{equation}
The case $\la c\ra <c^*$ can be studied similarly.
We would like to stress here, that Theorem \ref{thm:1} also allows us
to estimate $N|\la c\ra-c^*|^2$, but this estimate can depend on $\e$.

Now let us come back to (\ref{l4.4a}). In view
of (\ref{l4.6a}) for our choice of $A$
\begin{equation}\begin{array}{c}\ds{
\frac{A^2D^2}{8}-AN^{1/2}(\ti c-\la c\ra)\le \log 2\Rightarrow A\le
\frac{8N^{1/2}(\ti c-\la c\ra)+4D}{D^2}}\\
\ds{
\Rightarrow\max_U\phi_N^{(\mu)}(\ti c,U)\ge -2\frac{7(\ti
c-\la c\ra)^2+3(\la c\ra-c^*)^2}{D^2}-\frac{4}{N}}\ge
-14\frac{(\ti c-\la c\ra)^2}{D^2}-\frac{K_0}{N}
 \end{array}\label{l4.5a}\end{equation}
with some $N,\mu,\e$-independent $K_0$.

Let us take $L_1=8R_0$ and  $\ti c>\la c\ra+ L_1N^{-1/2}$. Consider
$\ti M(\ti c)\equiv N\max_U\phi_N^{(\mu)}(\la c\ra+2(\ti c-\la c\ra),U)$

If $\ti M(\ti c)< -4$, consider the sets 
\begin{equation}
\Pi_1\equiv\{(U,c):c>\ti c\},\quad 
\Pi_2\equiv\{(U,c):\la c\ra -L_1N^{-1/2}\le c\le\ti c\}.
\label{Pi_1,2}\end{equation}
Applying (\ref{p4.2}) and (\ref{l4.6b}), we get

\begin{equation}\begin{array}{c}\ds{
m(\Pi_1\cup \Pi_2)\ge \frac{3}{4},\,\,
m(\Lambda_{\ti M(\ti c)})\ge \frac{3}{4}}\\
\ds{
\Rightarrow m(\Lambda_{\ti M(\ti c)}\cap(\Pi_1\cup \Pi_2))\ge\frac{1}{2}
\ge m(\ov\Lambda_{\ti M(\ti c)}\cup(\ov\Pi_1\cap\ov\Pi_2))}\\
\ds{
\Rightarrow \la\theta(c-\ti c)\ra_{(U,c)}\ge
\frac{m(\Lambda_{\ti M(\ti c)}\cap\Pi_1)}
{m(\Lambda_{\ti M(\ti c)}\cap(\Pi_1\cup \Pi_2))+
m(\ov\Lambda_{\ti M(\ti c)}\cup(\ov\Pi_1\cap\ov\Pi_2))}}\\
\ds{
\ge \frac{m(\Lambda_{\ti M(\ti c)}\cap\Pi_1)}
{2(m(\Lambda_{\ti M(\ti c)}\cap\Pi_1)+m(\Lambda_{\ti M(\ti c)}\cap\Pi_2))}
\ge\frac{1}{2(1+e^{-\ti M(\ti c)}S_2S_1^{-1})},}
\end{array}\label{l4.5c}\end{equation}
where we denote by $S_{1,2}$ the Lebesgue measure of 
$\Lambda_{\ti M(\ti c)}\cap\Pi_{1,2}$, and use the fact that 
$0\ge N\phi_N^{(\mu)}(U,c)\ge \ti M(\ti c)$.

Consider the point $(\la c\ra +2(\ti c-\la c\ra), U_1)$, found from the
condition $ N\phi_N^{(\mu)}(\la c\ra +2(\ti c-\la c\ra), U_1)=\ti M(\ti c)$
and two points $(\ti c,U_2)$, $(\ti c,U_3)$ which belong to the 
boundary of $\Lambda_{\ti M(\ti c)}$. Since $\Lambda_{\ti M(\ti c)}$ is a
convex set, if we draw two straight lines through the first and the second 
and the first and the third points and
denote by $T$ the domain between these  lines, then
$T\cap\Pi_1\subset \Lambda_{\ti M(\ti c)}\cap\Pi_1$ and
$\Lambda_{\ti M(\ti c)}\cap\Pi_2\subset T\cap\Pi_2$. Therefore

\begin{equation}
\frac{S_1}{S_2}\ge \frac{(\ti c-\la c\ra)^2}
{(2(\ti c-\la c\ra)+L_1)^2-(\ti c-\la c\ra)^2}\ge \frac{1}{8}.
\label{l4.5d}\end{equation}
Thus, we derive from (\ref{l4.5c}):
\begin{equation}
\la\theta(c-\ti c\ra_{(U,c)}\ge\frac{e^{\ti M(\ti c)}}
{2e^{\ti M(\ti c)}+16}.
\label{l4.5e}\end{equation}
If $\ti M(\ti c)>-4$, let us chose $c_1>c^*$, which  satisfies
condition $N\max_U\phi_N^{(\mu)}(2c_1,U)=-4$ 
($c_1>\la c\ra+2(\ti c-\la c\ra)$). 
Replacing in the above consideration
$\Lambda_{\ti M(\ti c)}$ by $\Lambda_{-4}$, we finish the proof of the
first line of (\ref{l4}).

To prove the second line of (\ref{l4}) we choose any
$c_1>c^*+L_1N^{-1/2}$, which satisfies the
condition $N\max_U\phi_N^{(\mu)}(2c_1,U)<-4$, denote $d=2\e^{1/4}N^{-1/2}$
and  write similarly to (\ref{l4.5c})
\begin{equation}\begin{array}{c}\ds{
\la \chi_{c^*,c^*+ d}(c)\ra_{(U,c)}
\le\frac{m(\Lambda_{-4}\cap\Pi_{c^*,c^*+d})+
m(\ov\Lambda_{-4}\cap\Pi_{c^*,c^*+d})}
{m(\Lambda_{-4}\cap\ov\Pi_{c^*,c^*+d})}}\\
\ds{
\le\frac{5m(\Lambda_{-4}\cap\Pi_{c^*,c^*+d})}
{4m(\Lambda_{-4}\cap\ov\Pi_{c^*,c^*+d})} }\\
\ds{
\le\frac{5 e^4\ti S_2}{4\ti S_1}\le\frac{5e^4}{4}\frac{(c_1-c^*)^2-
(c_1-c^*-d)^2}{(c_1-c^*-d)^2}
\le \e^{1/4}C_3^*,}
\end{array}\label{l4.5f}\end{equation}
where we denote by $\ti S_{1,2}$ the Lebesgue measures of 
$\Lambda_{-4}\cap\ov\Pi_{c^*,c^*+d}$ and $\Lambda_{-4}\cap\Pi_{c^*,c^*+d}$
respectively. 
Now, using the first line of (\ref{l4.7}), we obtain
the second line of (\ref{l4}).  Lemma \ref{lem:4} is proven.

\smallskip

\no{\it Proof of Proposition \ref{pro:4}} 

\no Let us  introduce  new
variables $\rho\equiv\sqrt{(c-c^*)^2+(U-U^*)^2}$, 

\no $\varphi\equiv$
$\arcsin\frac{U-U^*}{\sqrt{(c-c^*)^2+(U-U^*)^2}}$. Then
$\phi_N^{(\mu)}(\rho,\varphi)$ for any $\varphi $ is a concave
function of $\rho$. Let $r(\varphi)$ be defined from the condition
$N\phi_N^{(\mu)}(r(\varphi),\varphi) =M$. Consider
$ \phi_{M}(\rho,\varphi)\equiv 
r^{-1}(\varphi) \cdot\phi_N^{(\mu)}(r(\varphi),\varphi) \rho$.
 Since
$\phi_N^{(\mu)}(\rho,\varphi)$ is concave, we obtain that
\begin{equation}\begin{array}{ll}
\phi_N^{(\mu)}(\rho,\varphi)\ge\phi_{M}(\rho,\varphi),&\ds{ 0\le
\rho\le r(\varphi) ,} \\
\phi_N^{(\mu)}(\rho,\varphi)\le\phi_{M}(\rho,\varphi),&\ds{ \rho\ge
r(\varphi)}.
\end{array}\label{p4.3}\end{equation}
Thus, denoting  by $R$ the l.h.s. of the first inequality in (\ref{p4.2}),
we get
 $$\begin{array}{c}\ds{
 R\le\frac{\int
d\varphi\int_{\rho>r(\varphi)}d\rho
\exp\{N\phi_N^{(\mu)}(\rho,\varphi)\}} {\int d\varphi
\int_{\rho<r(\varphi)}d\rho
\exp\{\phi_N^{(\mu)}(\rho,\varphi)\}}}\\
\ds{ \le \frac{\int
d\varphi\int_{\rho>r(\varphi)}d\rho
\exp\{N\phi_M(\rho,\varphi)\}} {\int d\varphi\int_{\rho<r(\varphi)}d\rho
\exp\{N\phi_M(\rho,\varphi)\}} \le
\frac{(1-M)e^{M}}{1-(1-M))e^{M}}\le \frac{1}{4}.}
\end{array}$$
For the second inequality in (\ref{p4.2}) the proof is the same.
 To obtain  (\ref{p4.1}) let us remark first that due to the choice
 of $A$ the function
 $\phi_{\ti c}(\rho,\varphi)\equiv \phi_N^{(\mu)}(\rho,\varphi)+N^{-1/2}A\rho\cos\varphi$  for any
 $\varphi$ is a concave function of $\rho$, whose derivative
 at the point $\rho=\rho_\varphi\equiv\ti c|\cos\varphi|^{-1}$ satisfies the condition
 $$
 \frac{d}{d\rho}\phi_{\ti c}(\rho_\varphi,\varphi)\le
 \frac{d}{d\rho}\phi_N^{(\mu)}(\rho_\varphi,\varphi)-\frac{1}{2}
\frac{\phi_N^{(\mu)}(\rho_\varphi,\varphi)}{\rho_\varphi}\le
\frac{1}{2}\frac{d}{d\rho}\phi_N^{(\mu)}(\rho_\varphi,\varphi).
 $$
Thus, for any $\varphi$ we can  write
 $$\begin{array}{c}
\ds{
\frac{\int_{\rho>\rho_\varphi}d\rho
e^{N\phi_N^{(\mu)}(\rho,\varphi)}e^{AN^{1/2}(\cos\varphi\rho-\ti c)}}
{\int_{\rho>\rho_\varphi}e^{N\phi_N^{(\mu)}(\rho,\varphi)}}
 \le
  \frac{|{d\over
d\rho}\phi_N^{(\mu)}(\rho_\varphi,\varphi)+
AN^{-1/2}\cos\varphi|^{-1}} {|{d\over
d\rho}\phi_N^{(\mu)}(\rho_\varphi,\varphi)|^{-1}}
\le 2.}
\end{array}$$
This inequality implies (\ref{p4.1}).

\smallskip

\no {\it Proof of Proposition \ref{pro:5}}. To prove 
Proposition \ref{pro:5} we use the method, developed 
in \cite{PST2}.
 Consider the function $g(A)$ defined by (\ref{l4.5}) and let us write
 the Taylor expansion up to the second order with respect to $t$ for
 $g(tA)$ ($t\in [o,1]$). Then
\begin{equation}\begin{array}{c}
R_N=A^2\int_0^1dt (1-t)g''(tA)dt-{1\over 2}A^2g''(0)\\
=A^3\int_0^1dt(1-t)\int_0^tdt_1N^{-3/2}\sum\xmi\la(\dot{\bJ},\dot{\bJ})\dot J_i
\ra_{\mu,t_1}\\
+A^2\int_0^1dt(1-t)N^{-1}\sum_{i\not=j}\xmi\xmj\la\dot J_i\dot J_j\ra_{\mu,t}
\equiv R_N^{(1)}+R_N^{(2)},
\end{array}\label{l4.8}\end{equation}
where we denote
$$
\la...\ra_{\mu,t}\equiv\frac{\lla(\dots)\exp\{tAN^{-1/2}(\bxm,\bJ)\}\rra_\mu}
{\lla\exp\{tAN^{-1/2}(\bxm,\bJ)\}\rra_\mu}.
$$
Let us estimate
\begin{equation}\begin{array}{c}\ds{
E\{(R_N^{(1)})^4\}\le A^{12}N^{-6} \int_0^1dt
\biggl(\sum_{i_1\not=i_2\not=i_3\not=i_4}E\{\xi^{(\mu)}_{i_1}
\xi^{(\mu)}_{i_2}\xi^{(\mu)}_{i_3}\xi^{(\mu)}_{i_4}}\\
\ds{
\la(\dot{\bJ},\dot{\bJ})\dot J_{i_1}\ra_{\mu,t}
\la(\dot{\bJ},\dot{\bJ})\dot J_{i_2}\ra_{\mu,t}
\la(\dot{\bJ},\dot{\bJ})\dot J_{i_3}\ra_{\mu,t}
\la(\dot{\bJ},\dot{\bJ})\dot J_{i_4}\ra_{\mu,t}\}}\\
\ds{
+6\sum_{i_1\not=i_2\not=i_3}E\{
\xi^{(\mu)}_{i_2}\xi^{(\mu)}_{i_3}
\la(\dot{\bJ},\dot{\bJ})\dot J_{i_1}\ra_{\mu,t}^2
\la(\dot{\bJ},\dot{\bJ})\dot J_{i_2}\ra_{\mu,t}
\la(\dot{\bJ},\dot{\bJ})\dot J_{i_3}\ra_{\mu,t}\}}\\
\ds{
3\sum_{i_1\not=i_2}E\{
\la(\dot{\bJ},\dot{\bJ})\dot J_{i_1}\ra_{\mu,t}^2
\la(\dot{\bJ},\dot{\bJ})\dot J_{i_2}\ra_{\mu,t}^2\}}\\
\ds{
4\sum_{i_1\not=i_2}E\{\xi^{(\mu)}_{i_1}
\xi^{(\mu)}_{i_2}
\la(\dot{\bJ},\dot{\bJ})\dot J_{i_1}\ra_{\mu,t}^3
\la(\dot{\bJ},\dot{\bJ})\dot J_{i_2}\ra_{\mu,t}\}+
\sum_{i_1}E\{
\la(\dot{\bJ},\dot{\bJ})\dot J_{i_1}\ra_{\mu,t}^4\}\biggr).}
\end{array}\label{l4.9}\end{equation}
Now, using the formula of integration by parts (\ref{ibp}), taking
into account that in our case
${\d\over\d\xmi}=Ath^{-1}N^{-1/2}{\d\over\d h_i}$,
and then using integrations by parts with respect to the Gaussian variable
$h_i$, one can substitute
\begin{equation}\begin{array}{c}
E\{\xmi\la\dots\ra_{t,\mu}\}\to
Ath^{-1}N^{-1/2}E\{ h_i\la\dots\ra_{t,\mu}\}\\
+N^{-3/2}A^3
O(E\{\la(\dot J_i)^2(\dots)\ra_{t,\mu}\}).
\end{array}\label{l4.9a}\end{equation}
Thus, for the first sum in (\ref{l4.9}), we obtain
\begin{equation}\begin{array}{c}
E\{\Sigma_1\}\le h^{-4}A^{16}N^{-8} \int_0^1dt
E\left\{\left(\sum_{i_1}h_{i_1}\la(\dot{\bJ},\dot{\bJ})\dot J_{i_1}\ra_{\mu,t}
\right)^4\right\}+O(A^{18}N^{-3})\\
\le  h^{-4}A^{16}N^{-2} \int_0^1dt
E\left\{\left(N^{-1}\sum_{i,j}h_ih_j\la\dot J_i\dot J_j\ra_{\mu,t}\right)^2
\la(N^{-1}(\dot{\bJ},\dot{\bJ}))^2\ra^2\right\}\\
\le
\const A^{16} N^{-2}.
\end{array}\label{l4.10}\end{equation}
Here to estimate the errors term in (\ref{l4.9a}) we use that,
according to Theorem \ref{thm:1} (see (\ref{t1.1})),
for any fixed $p$ $E\{\la\dot J_i^p\ra_{\mu,t} \}$  is bounded by
$N$-independent constant.

Other sums in the r.h.s. of (\ref{l4.9})  and $E\{(R_N^{(1)})^4\}$
can be estimated similarly to (\ref{l4.10}).

\smallskip

\no{\bf Acknowledgements.} Authors would like to thank Prof.A.D.Milka
for the fruitful discussion of the geometrical aspects of the problem.

\end{document}